\documentclass[12pt, draftclsnofoot, onecolumn]{IEEEtran}
\usepackage{mathtools}
\usepackage{tikz}
\usepackage{graphicx}
\usepackage[english]{babel}
\usepackage{epsfig}
\usepackage{graphics}
\usepackage{cite}
\usepackage{amsmath}
\usepackage{amssymb}
\usepackage{amsfonts}
\usepackage{setspace}
\usepackage{color}
\usepackage{setspace}
\usepackage{multirow}
\usepackage{tablefootnote}
\usepackage{dblfloatfix}
\usepackage{multicol}
\usepackage{algorithm,algorithmic}
\usepackage{subcaption}
\usepackage{bm, amsmath, amssymb}
\usepackage{amsfonts}
\usepackage {amssymb,amsmath}
\usepackage{cite}
\usepackage{epstopdf}
\usepackage{booktabs}
\usepackage{float}
\usepackage{cite}
\usepackage{wrapfig}
\graphicspath{ {W01Spaper_Images_26082018/} }
\usepackage{mathrsfs}
\usepackage{tikz}
\mathchardef\mhyphen="2D

\makeatletter
\newcommand{\removelatexerror}{\let\@latex@error\@gobble} 
\makeatother
\newcommand{\sm}[1]{{\normalsize{#1}}}

\newcommand{\ovl}[1]{\overline{#1}}
\newcommand{\mcal}[1]{\mathcal{#1}}
\newcommand{\mbf}[1]{\mathbf{#1}}
\newcommand{\mcalR}[1]{\mcal{R}}

\DeclareMathOperator{\N}{\mathbb{N}}
\DeclareMathOperator{\C}{\mathbb{C}}
\newcommand{\stonen}{\widetilde{1}}
\newcommand{\sttwon}{\widetilde{2}}
\newcommand{\stthrn}{\widetilde{3}}
\newcommand{\stnn}{\widetilde{\N}}

\DeclareGraphicsExtensions{.pdf,.png,.jpg}

\DeclareMathOperator{\E}{\mathbb{E}}

\DeclareMathOperator{\SSS}{\mcal{S}}
\DeclareMathOperator{\SR}{\mcal{R}}
\DeclareMathOperator{\SD}{\mcal{D}}
\DeclareMathOperator{\PD}{\mcal{P}}
\DeclareMathOperator{\U}{\textit{u}}

\IEEEoverridecommandlockouts

\usepackage{tikz}
\usepackage{textcomp}
\usepackage{hyperref}
\usepackage{lipsum}
\usepackage[stable]{footmisc}

\addto\captionsenglish{}
\usepackage[font=scriptsize]{caption}
\usepackage{mwe}

\newcommand\copyrighttext{%
	\footnotesize "This work has been submitted to an IEEE journal for possible publication. Copyright may be transferred without prior notice, after which this version may no longer be accessible"}
\newcommand\copyrightnotice{%
	\begin{tikzpicture}[remember picture,overlay]
	\node[anchor=south,yshift=760pt] at (current page.south) {{\parbox{\dimexpr\textwidth-\fboxsep-\fboxrule\relax}{\copyrighttext}}};
	\end{tikzpicture}%
}

\newsavebox{\ieeealgbox}

\begin{document}
	
	\title{Framework for Discrete Rate Transmission in Buffer-Aided Underlay CRN With Direct Path}

\author{Bhupendra Kumar,~\IEEEmembership{Student Member,~IEEE,} and Shankar Prakriya,~\IEEEmembership{Senior Member,~IEEE}	
	\thanks{This work was supported by Information Technology Research Academy through sponsored project ITRA/15(63)/Mobile/MBSSCRN/01.}
	\thanks{Bhupendra Kumar is with the Bharti School of Telecom Tech. and Management, IIT Delhi (e-mail: bkumar0810@gmail.com). S. Prakriya is with the Department of Electrical Engineering, IIT Delhi, New Delhi 110016, India (e-mail: shankar@ee.iitd.ac.in).}
}

	\maketitle
	\copyrightnotice
\vspace{-1.5 cm}
\begin{abstract}
	
	\par In this paper,   a buffered decode and forward (DF) relay based three-node underlay cooperative cognitive relay network (CRN) is considered with a direct path to the destination. The source and the relay use multiple rates, and joint rate and link selection is performed to maximize throughput.  Optimum link and rate selection rules are evolved that ensure buffer stability, and expressions are derived for the throughput assuming peak power and peak interference constraints on the transmit power of the secondary nodes.  The expressions are written in a manner that yields useful insights on buffer stability and role of the direct link on performance. A scheme in which the direct link signal is combined with the relayed signal is also considered, and it is demonstrated that it offers additional improvement in performance only in some scenarios. Computer simulations have been presented to verify the accuracy of derived expressions.
\end{abstract}

\section{Introduction}	


\par Due to rapid increase in  demand for data intensive applications and services,  and profileration of wireless devices, the wireless industry today faces an acute spectrum shortage.   Cognitive radio technologies are seen to be a solution to this shortage. Underlay type of cognitive radios, in which the  transmit powers of secondary nodes is constrained to ensure that  interference to  the primary licensed users is below a certain interference temperature limit \cite{Xing2007,Goldsmith2009},  have shown  great potential in increasing spectrum utilization efficiencies. Due to these constraints on the transmit powers,  relays are often needed in the secondary network to increase range and reliability \cite{Ding2008}\cite{Hu2016}. 
\par Although they incur a loss in performance as compared to full-duplex relays \cite{james2011}, half-duplex relays are preferred in many situations  because of their simplicity.   One  option to overcome the loss due to half-duplex relays is to utilize rate selection, which requires channel knowledge at the transmitter\cite{Issariyakul2006}.  In addition, when the direct link between the source and the destination is not shadowed, combining the direct and relayed signals improves performance and harnesses diversity gain in cooperative links.  Yet another option to harness diversity gain is to use link selection, which requires the incorporation of a data buffer at the relay \cite{Zlatanov2014,Nomikos2015}.  Use of data buffers in relays provides some degree of freedom in scheduling links degraded by fading, and increases throughput. For this reason, buffer-aided relaying has been investigated in different scenarios extensively (relay-selection \cite{Teh2015,Raja2018}, multi-hop \cite{Tian2016,Yang2018}, two-way relaying \cite{Jamali2014_1,Jamali2014_2}, MIMO systems\cite{Tang2018},  energy harvesting\cite{Liu2018}, physical layer security\cite{Liao2018,Wang2018}, NOMA\cite{Zhang2017,Cao2018}, full-duplex relays \cite{Razlighi2018,Nomikos2018} and CRN \cite{Shaqfeh2015,Kulkarni2017} etc.). Analysis of performance of buffered relays in underlay CRN has been carried out for half and full-duplex relays in  \cite{Darabi2015,Darabi2017} and \cite{Kumar2018} respectively.

\subsection*{Motivation and Contributions}
\par Due to the interference constraints, the link signal to noise ratios (SNRs) in underlay  cognitive radio network (CRN) have large variance. For this reason, use of a buffer-aided relay with link adaptation is appealing in CRN \cite{Kumar2018}\cite{Zlatanov2013_1,Zlatanov2013_2,Wicke2017}. For the same reason, use of rate selection is well motivated in CRN, and we investigate this aspect here. As noted already, due to power constraints, the nodes in underlay networks are relatively close to each other for acceptable quality of service (QoS).  Taking the direct channel into consideration is therefore important in underlay CRNs.   In this work, we consider the direct channel, and perform joint rate and link selection with buffered relays in a two-hop underlay cognitive network.   The major contributions of our work are as follows:
\begin{itemize}
	
	\item We provide a general framework for discrete-rate transmission in underlay cognitive relay networks with a direct path. We first develop the joint rate and link-selection protocol and analyze the prerequisite for buffer stability.
	
	\item We then rewrite the throughput in a manner that provides deep insights into performance\footnote{Delay analysis is clearly of interest, but it is not included here due to paucity of space. It will be studied separately.} and buffer stability.  
	
	\item We utilize the expressions to analyze throughput performance of  two schemes. In the first one,  joint link and rate selection is performed amongst the three links. In the second scheme, the  relay and the source signal using  OSTBC based on the Alamouti code whenever the R-D link is selected. To enable analysis of performance, expressions are derived for joint complementary commutative distribution function (CCDF) of instantaneous SNRs of the links for both the schemes. Note that expressions for performance of the traditional non-cognitive cooperative network follow as a special case.
\end{itemize}

\section{System Model}\label{sec:SysMod}
\par We consider a dual-hop underlay cooperative CRN  as depicted in Fig.\ref{fig:sysmod1} in which the primary network consists of a primary source (not depicted in the figure), and a primary destination ($\PD$). The secondary or unlicensed network consists of the secondary source ($\SSS$), the secondary destination ($\SD$), together with a half-duplex (HD) decode and forward (DF) buffer-aided secondary relay ($\SR$). All these nodes are assumed to possess a single antenna. 
\subsubsection*{Channel Model}
In this paper links 1, 2 and 3 refer to  $\SSS-\SR$, $\SR-\SD$, and $\SSS-\SD$ channels respectively.  The links are of fading type  with coefficients $h_i(n)$, $i=1,2,3$. The interference channels from $\SSS$ and $\SR$ to $\PD$ are denoted by $g_1(n)$ and $g_2(n)$ respectively. We will find it convenient to define $g_3(n)=g_1(n)$. We assume Rayleigh fading channels so that $h_i(n)\sim{\cal CN}\left(0,\Omega_{h_{i}}\right)$,  and $g_i(n)\sim {\cal CN}\left(0,\Omega_{g_{i}}\right)$, $i=1,2,3$.
Denote by ${\mathbb I}_{p}$ the interference temperature limit (ITL) imposed by the primary network, and by \sm{$\mathbb{P}_{max}$} the maximum transmit power  at $\SSS$ and $\SR$.
We denote by \sm{$\gamma_{i}(n)$} the instantaneous SNR of link \sm{$i,\,\forall i\in\{1,2,3\}$}. Let $\gamma_{max}= \mathbb{P}_{max}/N_{o}$, and  $\gamma_{p}={\mathbb{I}}_{p}/N_{o}$, where $N_o$ is the  power spectral density of additive white Gaussian noise samples.  For underlay cognitive radio with peak transmit power (PTP) and peak interference power (PIP) constraints, $\gamma_i(n)$ is given by:
\begin{IEEEeqnarray}{rCl}\label{eqn:InsSNR}
	\hspace{-0.5cm}\gamma_{i}(n) &=& \min\left\{\gamma_{max},\frac{\mcal{\gamma}_{p}}{|g_{i}(n)|^2}\right\}|h_{i}(n)|^2.  \label{eqn:link_snrs}
\end{IEEEeqnarray}
We assume quasi-static Rayleigh fading channels with path-loss exponent $\alpha$. Hence, $\Omega_{h_{i}}=d_{i}^{-\alpha}$ and $\Omega_{g_{i}}=d_{ip}^{-\alpha}$, where $d_{i}$ and $d_{ip}$ respectively denote (for link-$i$) the distances between nodes in the main and interference link . The probability $p_{i}$, $i\in \{1,2\}$,   that  the peak interference ($\mathbb{P}_{max}|g_{i}(n)|^{2}$) at $\PD$ is greater than ${\mathbb I}_{p}$ when transmit power $\mathbb{P}_{max}$ is used, is given by \cite{Kumar2018}:
\begin{IEEEeqnarray}{rCl}\label{eqn:p_i}
p_{i}&=&\Pr\left\{\gamma_{max}>\dfrac{\gamma_{p}}{|g_{i}(n)|^2}\right\}= e^{-{\mu_{i}}/{\lambda_{i}}},
\end{IEEEeqnarray}
where $\lambda_{i}=\gamma_{max}\,\Omega_{h_{i}}$ and $\mu_{i}=\frac{\gamma_{p}\Omega_{h_{i}}}{\Omega_{g_{i}}}$ represent the average transmit SNRs when $\SSS$ and  $\SR$ (respectively) transmit with powers $\mathbb{P}_{max}$ and $\mathbb{I}_p/\Omega_{g_{i}}$.   We note once again that $d_{3p}=d_{1p}$ hence $g_{3}(n)=g_{1}(n)$ and $p_{3}=p_{1}$. These notations are used for maintaining consistency in formulating the problem.
\subsubsection*{Rate Set}	
\par Joint link and rate selection is performed in this paper. We assume that $\SSS$ and $\SR$ use capacity achieving codewords of single time slot and pick transmission rate $R_{i}^{k_{i}}$ when the $i^{th}$ link is selected. Let $R_i^{[0,1,\ldots,K_i]}=[R_i^0,R_i^1,\ldots,R_i^{K_i}]$ denote the rate vector with rates arranged in increasing order so that:
\begin{IEEEeqnarray}{rCl}\label{eqn:rate_set}
	\begin{array}{l}
		\text{Rate set}\ \{R_{1}^{[0,1,...k_{1}...K_{1}]}\}\equiv\text{SNR threshold set}\  \{\gamma_{1}^{[0,1,...k_{1}...K_{1}]}\} \text{ for link-1 } (\SSS-\SR \text{link}),\\
		\text{Rate set}\ \{R_{2}^{[0,1,...k_{2}...K_{2}]}\} \equiv\text{SNR threshold set}\  \{\gamma_{2}^{[0,1,...k_{2}...K_{2}]}\} \text{ for link-2 } (\SR-\SD \text{link}),\\
		\text{Rate set}\ \{R_{3}^{[0,1,...k_{3}...K_{3}]}\} \equiv\text{SNR threshold set}\  \{\gamma_{3}^{[0,1,...k_{3}...K_{3}]}\} \text{ for link-3 } (\SSS-\SD \text{link}),
	\end{array}
\end{IEEEeqnarray}
where $\gamma_{i}^{k_{i}}$ is the SNR threshold for the rate $R_{i}^{k_{i}}$, which is defined as $\gamma_{i}^{k_{i}}=2^{R_{i}^{k_{i}}}-1$. Note that the rate set for the $\SSS-\SR$ and $\SSS-\SD$ links are identical so that  $K_1=K_3$, and we choose a different index  $k_{3}$  for the third link $\SSS-\SD$ only for ease of exposition. Also note that initial rate is zero for every rate-set, i.e. \sm{$R_{i}^{0}=0$ hence $\gamma_{i}^{0}=0$ for  $i\in\{1,2,3\}$}.

\begin{figure}[t]
	\begin{center}
		\includegraphics[width=69mm]{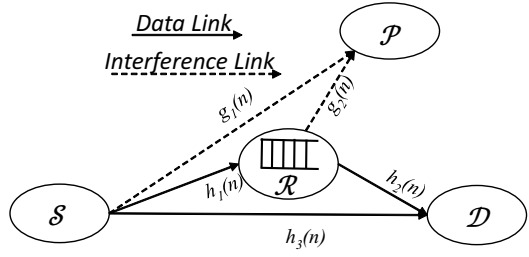}
		\caption{Three Node cognitive buffer-aided relay network.}
		\label{fig:sysmod1}
	\end{center}
	\vspace{-0.75cm}
\end{figure}

\subsubsection*{Link Selection Variables}
\par  We will find it useful to define indicator functions that specify if a particular rate is selected for a link. Specifically, we define the  $2(K_{1}+1)+(K_{2}+1)$ indicator functions \sm{$\U_{i}^{k_{i}}(n)$} as follows:
\begin{IEEEeqnarray}{rCl}\label{eqn:coupled_elements}
	\U_{i}^{k_{i}}(n)\hspace{-0.1cm}=\hspace{-0.1cm}\Bigg\{
	\begin{array}{ll}
		\hspace{-0.15cm}1 \,\,  \text{Rate $R_{i}^{k_i}$ is selected  }\\
		\hspace{-0.15cm}0 \,\, \text{otherwise}.
	\end{array} i=\{1,2,3\}
\end{IEEEeqnarray}
Clearly, we can generate the binary link selection variables $\U_{i}(n)$  as follows:
\begin{IEEEeqnarray}{rCl}
	\begin{array}{lll}
		\U_{i}(n)=\sum\limits_{k_{i}=0}^{K_{i}}\U_{i}^{k_{i}}(n).\quad  i\in\{1,2,3\} 
	\end{array}
\end{IEEEeqnarray}
For selecting any one of the three links for transmission, we define a link selection vector as $\mbf{u}(n)=[\U_{1}(n),\U_{2}(n),\U_{3}(n)]$. Note that $\displaystyle\sum_{i=1}^{3}\U_i(n)=1$ since only one rate corresponding to one link is selected.
\subsubsection*{Channel State Parameter and Set of Permissible Rates}
For selection of a link, and the rate to be used on it, we clearly require  information on whether  the channel is in outage for that rate. These indicator variables depend on the signalling scheme. We consider two signalling schemes in this paper. For links-1 and 3, we define these indicator functions for both the signalling schemes as follows:
\begin{IEEEeqnarray}{rCl}\label{eqn:ChanIndi_1_3}
\begin{array}{llllll}
I_{u_{1}}^{k_{1}}(n)\hspace{-0.2cm}&=&\hspace{-0.2cm}\Bigg\{
\begin{array}{ll}
1\quad  \text{if } R_{1}^{k_{1}}(n)\leq\,\log_{2}(1+\gamma_{1}(n))\\
0\quad \text{otherwise,}
\end{array} &\hspace{-0.4cm}
I_{u_{3}}^{k_{3}}(n)\hspace{-0.2cm}&=&\hspace{-0.2cm}\Bigg\{
\begin{array}{ll}
1\quad  \text{if } R_{3}^{k_{3}}(n)\leq\,\log_{2}(1+\gamma_{3}(n))\\
0\quad \text{otherwise.}
\end{array}
\end{array}
\end{IEEEeqnarray}
It is apparent that $I_{\U_{i}}^{k_{i}}(n)R_{i}^{k_{i}}(n),\, k_{i}=0...K_{i}$ can be thought of as the decodable rate set for link-$i$ ($i=1,3$), and its maximum value $R_{i}^{k_{i}^{*}}(n)=\displaystyle \max_{k_i=1,\ldots,K_i} (I_{\U_{i}}^{k_{i}}(n) R^{k_i}_{i}(n))$ is the best feasible rate for that link. For link-2, the defintion of $I_{\U_{2}}^{k_{2}}(n)$ depends on the signalling scheme used. In scheme-1, when link-$2$ is selected (in a manner to be discussed later), the relay transmits with rate $R_{2}^{k_{2}^{*}}(n)=\displaystyle \max_{k_2=1,\ldots,K_2} (I_{\U_{2}}^{k_{2}}(n) R^{k_2}_{2}(n))$. In scheme-2, both source and relay use the Alamouti orthogonal space-time block code (OSTBC) to transmit a packet to the destination when link-2 is selected, exploiting the fact that the same packets are also present at the source. Clearly, this scheme results in higher SNR at the destination.
\par In scheme-1,  the definition follows that used for links 1 and 3 so that:
\begin{IEEEeqnarray}{rCl}\label{eqn:ChanIndi_2_s1}
	I_{u_{2}}^{k_{2}}(n)=\Bigg\{
	\begin{array}{ll}
		1\ \   \text{if } R_{2}^{k_{2}}(n)\leq\,\log_{2}(1+\gamma_{2}(n)),\\
		0\ \  \text{otherwise.}
	\end{array}
\end{IEEEeqnarray}
\par In scheme-2, due to the distributed Alamouti coding, we have:
\begin{IEEEeqnarray}{rCl}\label{eqn:ChanIndi_2_s2}
	I_{u_{2}}^{k_{2}}(n)=\Bigg\{
	\begin{array}{ll}
		1\ \   \text{if } R_{2}^{k_{2}}(n)\leq\,\log_{2}(1+\gamma_{2}(n)+\gamma_{3}(n)),\\
		0\ \  \text{otherwise.}
	\end{array}
\end{IEEEeqnarray}
To facilitate scheme-2, we assume that the rate sets at the source and relay are identical,  and that the source tracks the buffer-content of the relay using a feedback link. Clearly, $R_{2}^{k_{2}^{*}}(n)=\displaystyle \max_{k_2=1,\ldots,K_2} (I_{\U_{2}}^{k_{2}}(n) R^{k_2}_{2}(n))$.  Furthermore, since the rate sets at the source and relay are identical for scheme-2, it is clear by comparing (\ref{eqn:ChanIndi_1_3}) and (\ref{eqn:ChanIndi_2_s2}) that only rates $R^{k_2}_{2}\geq R^{k_3}_{3}$ (or equivalently index $k_{2}\geq k_{3}$) are  permissible for scheme-2. On the contrary, scheme-1 has no such restriction due to the independence of its outage indicator functions.
\subsubsection*{Implementation of the Protocol}
$\SSS$ and $\SR$ estimate $|g_1(n)|^{2}$ and $g_2(n)|^{2}$ by observing reverse channel of the primary network, or using dedicated pilots transmitted by the primary receiver ${\cal P}$.  A pilot transmitted by $\SD$ enables $\SSS$ and $\SR$ to estimate $|h_{3}(n)|^{2}$ and $|h_{2}(n)|^{2}$ respectively.  Similarly, a pilot transmitted by $\SR$ enables $\SSS$ to estimate $|h_1(n)|^{2}$. We therefore assume that $\SSS$ has knowledge of $|g_1(n)|^{2}$, $|h_{1}(n)|^{2}$ and $h_3(n)|^{2}$, and that $\SR$ posseses knowledge $|g_2(n)|^{2}$ and $h_2(n)|^{2}$.  Indices $k_{1}^{*},\, k_{2}^{*}$ and $k_{3}^{*}$ of best rates of each link are selected  and passed on   to a control unit, which then determines the link selection that maximizes throughput.
\section{Implementation of Link Selection}\label{sec:ProbFormu}
\par In this section, we first formulate the throughput maximization problem and determine the optimal scheduling of reception and transmission. 
\subsubsection*{Throughput Maximization}\label{sec:Throu}
The average link-rate of link-$i$ over $N$ transmissions is given by:
\begin{IEEEeqnarray}{rCl}{\label{eqn:taufin}}
	\begin{array}{lll}
		\tau_{i} &=&\dfrac{1}{N} \sum\limits_{n=1}^{N}\sum\limits_{k_{i}=0}^{K_{i}}\U_{i}^{k_{i}}(n)I_{\U_{i}}^{k_{i}}(n)R_{i}^{k_{i}}(n). \quad \forall i\in\{1,2,3\}				
	\end{array}
\end{IEEEeqnarray}
\par The maximum feasible rate $R_{i}^{k_{i}^{*}}(n)$ in link-$i$  is given by:
\begin{IEEEeqnarray}{rCl}\label{eqn:maxfearate}
	\begin{array}{lll}
		\hspace{-0.5cm}R_{i}^{k_{i}^{*}}(n)=\underset{k_{i}}{\max}(I_{\U_{i}}^{k_{i}}(n)R_{i}^{k_{i}}(n)).
	\end{array}
\end{IEEEeqnarray}
We note that the system throughput $\tau_t$ needs to be maximized ($\tau_t=\tau_2+\tau_3$) by suitable selection of the signalling rates and the binary link selection variables $\U_i(n)$ in an optimal fashion while ensuring buffer stability ($\tau_1\leq \tau_2$). The optimization problem can be written as\footnote{We assume finite $N$ initially  as the  link and rate selection policies  remain the same for both finite and infinite $N$.}:
\begin{IEEEeqnarray}{rCl}
	\begin{array}{lll}\label{org_constraint}
		\hspace{2in}	\underset{\mbf{\mbf{\U}}(n)}{\max} \quad {\tau}_{t}={\tau}_{2}+{\tau}_{3}\\
		\text{ s.t. \hspace{1cm}   $\C_{0}$}: \tau_{1}\leq \tau_{2},\hspace{1cm}
		\text{ $\C_{1}$}: \U_{i}^{k_{i}}(n)(1-\U_{i}^{k_{i}}(n))=0,\hspace{1cm}  
		\text{ \hspace{.3cm}$\C_{2}$}: \sum\limits_{i=1}^{3}\sum\limits_{k_{i}=0}^{K_{i}}\U_{i}^{k_{i}}(n)=1.
	\end{array}
\end{IEEEeqnarray}
Since allowing the link selection variables to take values between $0$ and $1$ simplifies the problem but leads to the same solution \cite{Darabi2017}, we relax the binary constraint on them.   We note that for infinite-size buffers, using $\tau_1<\tau_2$ for buffer stability simply leads to loss in throughput. We therefore optimize so that $\tau_1=\tau_2$. The throughput maximization problem can then be re-written as follows:
\begin{IEEEeqnarray}{rCl}
	\begin{array}{lll}
		\hspace{1.5in}	\underset{\mbf{\mbf{\U}}(n)}{\max} \quad {\tau}_{t}={\tau}_{2}+{\tau}_{3}\hspace{2cm}
		\text{ s.t.  \hspace{1cm}  $\C_{0}$}: \tau_{1}=\tau_{2},\\
		\text{ \hspace{.0cm}$\C_{1a}$}: \U_{i}^{k_{i}}(n)\geq 0, \text{ \hspace{.3cm}$\C_{1b}$}: \U_{i}^{k_{i}}(n)\leq 1,
		\text{ \hspace{.3cm}$\C_{2a}$}: \sum\limits_{i=1}^{3}\sum\limits_{k_{i}=0}^{K_{i}}\U_{i}^{k_{i}}(n)\geq 0,
		\text{ \hspace{.3cm}$\C_{2b}$}: \sum\limits_{i=1}^{3}\sum\limits_{k_{i}=0}^{K_{i}}\U_{i}^{k_{i}}(n)\leq 1.
	\end{array}
\end{IEEEeqnarray}
where constraint $\C_{0}$ is required for buffer stability, and linear constraints $\C_{1a},\,\C_{1b},\,\C_{2a}$ and $\C_{2b}$ arise on relaxing the binary constraint $\C_{1}$ in (\ref{org_constraint}).		
\subsubsection*{Mode of Operation}
\par We first note that several modes of operation arise depending on which nodes are eligible to transmit with non-zero rate. We denote by $\N$ the mode when no link is selected to transmit (all can transmit only with rate $0$),  and by $\stnn$ the mode when all links can transmit at some (non-zero) rate. Similarly, mode $i$ arises when only link-$i$ can  transmit at a non-zero rate, and $\widetilde{i}$ implies that all links other than $i$ can transmit at non-zero rate. As there are a total of $3$ links with $2$ states (on-off) each, the number of modes is clearly $2^{3}=8$. 
We represent mode by $e$ where $e \in \{\N,1,2,3,\stonen,\sttwon,\stthrn,\stnn\}\equiv\{i,\N,\widetilde{i},\stnn\}$. Please note that for convenience we denote the union of more than one mode, e.g. $\{1\cup\sttwon\cup\stthrn\cup\stnn\}$, which means the union of mode $1$, $\sttwon,\,\stthrn,$ and $\stnn$, by $\{1,\sttwon,\stthrn,\stnn\}$.
\subsubsection*{Coin-toss Events}
\par In situations when multiple links can  transmit at a non-zero rate, the solution to the optimization problem (as discussed in what follows) invokes coin toss to select a link.  In mode $\widetilde{i}$, the discrete rate $R_{j\neq i}^{k_{j}}(n)$ for link-$j$ is chosen by coin toss event $C_{j\neq i}^{\widetilde{i}}(n)$, whose probability is given as:
\begin{IEEEeqnarray}{rCl}\label{eqn:CT_Prob1}
	\begin{array}{lll}
		P_{j\neq i}^{\widetilde{i}}=\Pr\{C_{j\neq i}^{\widetilde{i}}(n)=1\}.	\quad i\in\{1,2,3\}		
	\end{array}
\end{IEEEeqnarray}
For example in mode $\stthrn$, where either link-$1$ or $2$ can be selected, a choice is made between  $R_{1}^{k_{1}^{*}}(n)$ and $R_{2}^{k_{2}^{*}}(n)$ by coin toss events $C^{\stthrn}_{1}(n)$ or $C^{\stthrn}_{2}(n)$, whose probabilities are $P^{\stthrn}_{1}=\Pr\{C^{\stthrn}_{1}(n)=1\}$ and $P^{\stthrn}_{2}=\Pr\{C^{\stthrn}_{2}(n)=1\}=1-P^{\stthrn}_{1}=\ovl{P}^{\stthrn}_{1}$ respectively. Similarly in mode $\stnn$, the discrete rate $R_{i}^{k_{i}^{*}}(n)$ for link-$i$ is chosen with the coin toss event $C_{i}^{\stnn}(n)$, with probabilities $P_{i}^{\stnn}=\Pr\{C_{i}^{\stnn}(n)=1\}	\quad i\in\{1,2,3\}$.
		
\subsubsection*{Lagrangian dual function and variables}
		
We now use the method of Lagrangian to perform the optimization. For convenience, we drop the  time-index in $\U_{i}^{k_{i}}(n)$ and $R_{i}^{k_{i}}(n)$. 
Using Lagrange multipliers $\alpha_w$, $\beta_i^{k_i}$, $\tilde{\beta}^{k_{i}}_i$, $\tilde{\beta}_{\N}$ and $\beta_{\N}$, we can write the Lagrangian cost function ${\cal L}$ as: 
\begin{subequations}\label{eqn:Lag1}
\begin{IEEEeqnarray}{rCl}
	\mcal{L}&&= -\tau_{2}-\tau_{3}-\alpha_{w} (\tau_{1}-\tau_{2}) -\sum\limits_{n=1}^{N}\Bigg[\sum\limits_{i=1}^{3}\sum\limits_{k_{i}=0}^{K_{i}}\Big[\beta_{i}^{k_{i}}\Big\{1-\U_{i}^{k_{i}}\Big\}\nonumber\\
	&&\hspace{0.35cm}+\widetilde{\beta}_{i}^{k_{i}}\,\U_{i}^{k_{i}}\Big]-\widetilde{\beta}_{\N}\Big[1-\sum\limits_{i=1}^{3}\sum\limits_{k_{i}=0}^{K_{i}}\U_{i}^{k_{i}}\Big]-\beta_{\N}\sum\limits_{i=1}^{3}\sum\limits_{k_{i}=0}^{K_{i}}\U_{i}^{k_{i}},
\end{IEEEeqnarray}
which can also be written as follows:
\begin{IEEEeqnarray}{rCl}\label{lag_2}
	\mcal{L}&&= -\alpha_{w} \tau_{1} - (1-\alpha_{w})\tau_{2}-\tau_{3} -\sum\limits_{n=1}^{N}\Bigg[\sum\limits_{i=1}^{3}\sum\limits_{k_{i}=0}^{K_{i}}\beta_{i}^{k_{i}}+\widetilde{\beta}_{\N}\Bigg]\nonumber\\
	&&\hspace{0.35cm}+\sum\limits_{n=1}^{N}\sum\limits_{k_{i}=0}^{K_{i}}\U_{i}^{k_{i}}\Bigg[\sum\limits_{i=1}^{3}(\beta_{i}^{k_{i}}-\widetilde{\beta}_{i}^{k_{i}})+\widetilde{\beta}_{\N}-\beta_{\N}\Bigg].
\end{IEEEeqnarray}
\end{subequations}
It is clear from the above equation that a group of Lagrangian multipliers is coupled with parameter $\U_{i}^{k_{i}}$. Define  $\varUpsilon_{i}^{k_{i}}$  as follows:
\begin{IEEEeqnarray}{rCl}\label{eqn:varUpsilon}
	{\varUpsilon_{i}^{k_{i}}}
	&=& N\big[(\beta_{i}^{k_{i}}-\widetilde{\beta}_{i}^{k_{i}})+\widetilde{\beta}_{\N}-\beta_{\N}\big].
\end{IEEEeqnarray}
Substituting the values of $\tau_{1},\tau_{2}$ and $\tau_{3}$ from (\ref{eqn:taufin}) in the expression for $\mcal{L}$ in (\ref{lag_2}), and using (\ref{eqn:varUpsilon}), we get:

\begin{IEEEeqnarray}{rCl}\label{eqn:lagrangian}
	\mcal{L}&&=\hspace{-0.125cm}-\frac{1}{N} \sum\limits_{n=1}^{N}\Bigg[\alpha_{w}\underset{\tau_{1}}{\underbrace{\sum\limits_{k_{1}=0}^{K_{1}}\U_{1}^{k_{1}}I_{\U_{1}}^{k_{1}}R_{1}^{k_{1}}}}+(1-\alpha_{w})\underset{\tau_{2}}{\underbrace{\sum\limits_{k_{2}=0}^{K_{2}}\U_{2}^{k_{2}}I_{\U_{2}}^{k_{2}}R_{2}^{k_{2}}}}\nonumber\\
	&&\hspace{0.2cm}+\hspace{-0.125cm}\underset{\tau_{3}}{\underbrace{\sum\limits_{k_{3}=0}^{K_{3}}\U_{3}^{k_{3}}I_{\U_{3}}^{k_{3}}R_{3}^{k_{3}}}}\hspace{-0.1cm}+\sum\limits_{i=1}^{3}\sum\limits_{k_{i}=0}^{K_{i}}(N\beta_{i}^{k_{i}}\hspace{-0.125cm}-\U_{i}^{k_{i}}\varUpsilon_{i}^{k_{i}})\hspace{-0.1cm}-N\widetilde{\beta}_{\N}\Bigg].
\end{IEEEeqnarray}
We note that  \ref{eqn:varUpsilon} and (\ref{eqn:lagrangian}) play a crucial role in the development of throughput maximization protocol.
\subsubsection*{Optimal rule for throughput maximization}
\par Now, we next state the link selection policy.
\par \emph{{\bf\emph Theorem 1:} The choice of the link-$i$ for transmission is carried out according to  (\ref{eqn:opt_sol})\footnote{${\bf x}^{T}$ denotes the transpose of vector ${\bf x}$.}, which is expressed in terms of the maximality of the rate decision metrics $\varUpsilon_{i}^{k_i*}(n)$  given by:
\begin{IEEEeqnarray}{rCl}\label{eqn:decision_metrics}
	\begin{array}{lll}
		\varUpsilon_{1}^{k_{1}^{*}}(n)&=&\alpha_{w^{*}} R_{1}^{k_{1}^{*}}(n),\quad
		\varUpsilon_{2}^{k_{2}^{*}}(n)=(1-\alpha_{w^{*}})R_{2}^{k_{2}^{*}}(n),\quad
		\varUpsilon_{3}^{k_{3}^{*}}(n)=R_{3}^{k_{3}^{*}}(n),
	\end{array}
\end{IEEEeqnarray}
where there exist parameter $\alpha_{w^{*}}$, and associated coin-toss probabilities\footnote{We highlight the dependence of these probabilities on $\alpha_w^{*}$ by writing these probabilities as functions of $\alpha_w^{*}$.} $P_{j\neq i}^{\widetilde{i}}(\alpha_{w^{*}})$ and $P_{i}^{\stnn}(\alpha_{w^{*}})$ such that the system throughput is maximized.}
\begin{figure}[!h]
	\hrulefill
	\begin{IEEEeqnarray}{rCl}\label{eqn:opt_sol}
		\begin{array}{lll}
			\hspace{-0.35cm}\U(n)\hspace{-0.1cm}\equiv
			
			\hspace{-0.1cm}\left[
			\begin{array}{l l l}\hspace{-0.2cm}u_{1}(n)\hspace{-0.1cm}\\\hspace{-0.2cm}u_{2}(n)\hspace{-0.1cm}\\\hspace{-0.2cm}u_{3}(n)\hspace{-0.1cm}\end{array}\hspace{-0.1cm}\right]^{T}\hspace{-0.2cm}=\hspace{-0.1cm}	\left\{
			\begin{array}{l l l}
				\hspace{-0.2cm}[0\hspace{1cm} 0\hspace{1.9cm} 0] \hspace{-0.125in}&:&\hspace{-0.1in} \varUpsilon_{1}^{k_{1}^{*}}(n)= \varUpsilon_{2}^{k_{2}^{*}}(n)= \varUpsilon_{3}^{k_{3}^{*}}(n)=0\ (\text{mode } \N),\\
				
				\hspace{-0.2cm}[1\hspace{1cm} 0\hspace{1.9cm} 0] \hspace{-0.125in}&:&\hspace{-0.1in} \varUpsilon_{1}^{k_{1}^{*}}(n)>\max(\varUpsilon_{2}^{k_{2}^{*}}(n),\varUpsilon_{3}^{k_{3}^{*}}(n))\ (\text{mode } 1),\\
				
				\hspace{-0.2cm}[0\hspace{1cm} 1\hspace{1.9cm} 0] \hspace{-0.125in}&:&\hspace{-0.1in} \varUpsilon_{2}^{k_{2}^{*}}(n)>\max(\varUpsilon_{3}^{k_{3}^{*}}(n),\varUpsilon_{1}^{k_{1}^{*}}(n))\ (\text{mode } 2),\\
				
				\hspace{-0.2cm}[0\hspace{1cm} 0\hspace{1.9cm} 1] \hspace{-0.125in}&:&\hspace{-0.1in} \varUpsilon_{3}^{k_{3}^{*}}(n)>\max(\varUpsilon_{1}^{k_{1}^{*}}(n),\varUpsilon_{2}^{k_{2}^{*}}(n))\ (\text{mode } 3),\\
				
				\hspace{-0.2cm}[0\hspace{1cm} C^{\stonen}_{2}(n) \hspace{.58cm}  {C^{\stonen}_{3}}(n)] \hspace{-0.125in}&:&\hspace{-0.1in} \varUpsilon_{2}^{k_{2}^{*}}(n)= \varUpsilon_{3}^{k_{3}^{*}}(n)> \varUpsilon_{1}^{k_{1}^{*}}(n)\hspace{1.cm} (\text{mode } \stonen),\\
				
				\hspace{-0.2cm}[C^{\sttwon}_{1}(n)\hspace{.5cm} 0 \hspace{1.25cm} C^{\sttwon}_{3}(n)]\hspace{-0.125in}&:&\hspace{-0.1in} \varUpsilon_{1}^{k_{1}^{*}}(n)= \varUpsilon_{3}^{k_{3}^{*}}(n)> \varUpsilon_{2}^{k_{2}^{*}}(n)\hspace{1.cm}(\text{mode } \sttwon),\\
				
				\hspace{-0.2cm}[C_{1}^{\stthrn}(n) \hspace{.48cm} C_{2}^{\stthrn}(n)\hspace{1.28cm} 0] \hspace{-0.125in}&:&\hspace{-0.1in}  \varUpsilon_{1}^{k_{1}^{*}}(n)= \varUpsilon_{2}^{k_{2}^{*}}(n)> \varUpsilon_{3}^{k_{3}^{*}}(n)\hspace{1.cm} (\text{mode } \stthrn),\\
				
				\hspace{-0.2cm}[C_{1}^{\stnn}(n)\hspace{.325cm} C_{2}^{\stnn}(n)\hspace{.45cm} C_{3}^{\stnn}(n)] \hspace{-0.125in}&:&\hspace{-0.1in} \varUpsilon_{1}^{k_{1}^{*}}(n)= \varUpsilon_{2}^{k_{2}^{*}}(n)= \varUpsilon_{3}^{k_{3}^{*}}(n)>0\ (\text{mode } \stnn).
			\end{array}
			\hspace{-0.25cm}\right.
		\end{array}
	\end{IEEEeqnarray}
	\hrulefill
\end{figure}
\par \noindent \emph{Proof:}
We first make the following observations about the optimization's conditions:	
\par 1) Dual Feasibility Condition: All the Lagrange multipliers for the inequality constraints have to be non-negative, i.e. $\widetilde{\beta}_{\N},\beta_{\N}\geq0$ and  $\beta_{i}^{k_{i}},\widetilde{\beta}_{i}^{k_{i}}\geq0,\,\forall i=\{1,2,3\}$. Further,  $0\leq \alpha_{w}\leq 1$.
\par 2) Complementary Slackness Condition: If an inequality is inactive, i.e. the optimal solution is in the interior of the set, the corresponding Lagrangian multiplier is zero. Therefore for $i=\{1,2,3\}$, we obtain:
\begin{IEEEeqnarray*}{rCl}
	\begin{array}{lll}
		\beta_{i}^{k_{i}}\Big[1-\U_{i}^{k_{i}}\Big]=0,\quad \widetilde{\beta}_{i}^{k_{i}}\,\U_{i}^{k_{i}}=0,\quad  	\widetilde{\beta}_{\N}\Big[1-\sum\limits_{i=1}^{3}\sum\limits_{k_{i}=0}^{K_{i}}\U_{i}^{k_{i}}\Big]=0,\quad \beta_{\N}\sum\limits_{i=1}^{3}\sum\limits_{k_{i}=0}^{K_{i}}\U_{i}^{k_{i}}=0.
	\end{array}
\end{IEEEeqnarray*}
\par 3) Stationarity w.r.t. primal variables:	
According to the stationary condition, differentiation w.r.t. to sub-primal variables $\U_{1}^{k_{1}},\U_{2}^{k_{2}}$ and $\U_{3}^{k_{3}}$ should be zero. Hence, we get:
\begin{IEEEeqnarray*}{rCl}
	\frac{\partial\mcal{L}}{\partial \U_{1}^{k_{1}}}
	=\varUpsilon_{1}^{k_{1}}- \alpha_{w} I_{\U_{1}}^{k_{1}}R_{1}^{k_{1}}=0,\,\,
	\frac{\partial\mcal{L}}{\partial \U_{2}^{k_{2}}}
	=\varUpsilon_{2}^{k_{2}}- (1-\alpha_{w}) I_{\U_{2}}^{k_{2}}R_{2}^{k_{2}}=0,\,\,
	\frac{\partial\mcal{L}}{\partial \U_{3}^{k_{3}}}
	=\varUpsilon_{3}^{k_{3}}- I_{\U_{3}}^{k_{3}}R_{3}^{k_{3}}=0.
\end{IEEEeqnarray*}
After solving the above stationary conditions, we get:
\begin{IEEEeqnarray}{rCl}\label{eqn:varUpsilon2}
	\begin{array}{lll}
		\varUpsilon_{1}^{k_{1}}
		=\alpha_{w} I_{\U_{1}}^{k_{1}}R_{1}^{k_{1}}, \hspace{2cm} {\varUpsilon_{2}^{k_{2}}}= (1-\alpha_{w})I_{\U_{2}}^{k_{2}}R_{2}^{k_{2}},
		\hspace{2.0cm}\varUpsilon_{3}^{k_{3}}= I_{\U_{3}}^{k_{3}}R_{3}^{k_{3}}.
	\end{array}	
\end{IEEEeqnarray}
Table-\ref{tab:rate_eval} lists the rate decision metrics for the case when either silence occurs, or one of the links is selected for transmission. It is accomplished by finding whether the relevant multipliers are active or not (using complementary slackness condition), and then using the multipliers in the stationarity condition.
\begin{table}[h]
	\normalsize 
	\caption{Rate selection-metrics for different link-selections}
	\normalsize 

	\renewcommand{\arraystretch}{1.2}
	\label{tab:lsp} 
	\begin{minipage}{\textwidth}
		\centering
		\resizebox{\textwidth}{!}{\begin{tabular}{|p{5cm}|p{12.cm}|}
			\hline\hline
			$\U_{i}(n)=0,\,\forall i\in\{1,2,3\}$\vspace{0.1cm} 
			& $\U_{i^{*}}(n)=1,\,\U_{i\neq i^{*}}(n)=0$ \\\hline
			
			\vspace{-0.2cm}$\beta_{i}^{k_{i}}=\widetilde{\beta}_{\N}=0,\,\forall i\in\{1,2,3\},$
			&\vspace{-0.2cm} $\widetilde{\beta}_{i^{*}}^{k_{i}^{*}}=\beta_{i\neq i^{*}}^{k_{i}}=\beta_{\N}=0,$\,\,\, $\beta_{i^{*}}^{k_{i}^{*}},\hspace{0.28cm}\widetilde{\beta}_{i\neq i^{*}}^{k_{i}},\hspace{0.28cm}\widetilde{\beta}_{\N}>0,$ \\
					
			\vspace{-0.2cm}$\widetilde{\beta}_{i}^{k_{i}},\hspace{0.28cm}\beta_{\N}>0,\,\forall i\in\{1,2,3\},$
			&\vspace{-0.2cm} $\varUpsilon_{i^{*}}^{k_{i}^{*}}=N(+\beta_{i^{*}}^{k_{i}^{*}}+\widetilde{\beta}_{\N})>0,$   \\
					
			\vspace{-0.2cm}$\varUpsilon_{i}^{k_{i}}=N(-\widetilde{\beta}_{i}^{k_{i}}-\beta_{\N})<0.$
			& \vspace{-0.2cm} $\varUpsilon_{i^{*}}^{k_{i}\neq k_{i}^{*}}=N(-\widetilde{\beta}_{i^{*}}^{k_{i}\neq k_{i}^{*}}+\widetilde{\beta}_{\N})<\varUpsilon_{i^{*}}^{k_{i}^{*}},\,\,\,$ $\varUpsilon_{i\neq i^{*}}^{k_{i}}=N(-\widetilde{\beta}_{i\neq i^{*}}^{k_{i}}+\widetilde{\beta}_{\N})<\varUpsilon_{i^{*}}^{k_{i}^{*}}.$\\\hline
		\end{tabular}}
		\label{tab:rate_eval} 
	\end{minipage}
\end{table}
\par Whenever silence occurs, $u_{i}^{k_{i}}=0$, therefore it is clear from the Table-\ref{tab:rate_eval} that $\varUpsilon_{i}^{k_{i}}<0$, $k_i\neq 0$.  Since $0\leq \alpha_{w}\leq 1$,  it is clear from (\ref{eqn:varUpsilon2}) that silence occurs when $\varUpsilon_{i}^{0}=\max_{k_i} (\varUpsilon_{i}^{k_{i}})=0 \Rightarrow R_{i}^{k_{i}^{*}}=0,\,\forall i.$ Whenever one of the link $i=i^{*}$ transmits, $u_{i^{*}}^{k_{i}^{*}}=1$, it is again clear from the Table-\ref{tab:rate_eval} that $\varUpsilon_{i^{*}}^{k_{i}^{*}}>0$ and $\varUpsilon_{i^{*}}^{k_{i}^{*}}>\varUpsilon_{i^{*}}^{k_{i}\neq k_{i}^{*}}$ and $\varUpsilon_{i^{*}}^{k_{i}^{*}}>\varUpsilon_{i\neq i^{*}}^{k_{i}}$. Hence in order to get optimum system throughput, we take the maximum of $\varUpsilon_{i}^{k_{i}}$ per time slot, which is expressed in terms of $R_{i}^{k_{i}^{*}}$ in (\ref{eqn:varUpsilon2}). Now, variable $\alpha_{w}$ adjusts the   selection of links and rates. If for some $\alpha_{w}$, the value of two or more decision metrics are the same, the link $i^{*}$ among these is chosen for transmission, i.e. $u_{i*}(n)= 1$, based on the relevant coin-toss probabilities.\ \ $\blacksquare$
\par It is clear that given $\alpha_w$, the maximal feasible rate of either link $1,\,2$ or $3$ is chosen if the corresponding decision-metric is greater than the decision-metric of the maximal feasible rate of other links (when there is equality, a coin toss is used to select a link). When  $\varUpsilon_{i}^{k_i*}$ is maximum, then its corresponding $u_{i}^{k_{i}^{*}}$ is set to unity (which also sets $\beta_{i^{*}}^{k_{i}^{*}},\,\widetilde{\beta}_{\N}$ and resets $\beta_{\N},\,\widetilde{\beta}_{i^{*}}^{k_{i}^{*}}$). The choice of  $\alpha_{w^{*}}$ and the coin toss probabilities depend on the channel statistics and is discussed in the following sections.
\section{Performance Evaluation}\label{sec:PerAna}
\par In this section, we express the throughput in terms of the rates obtained in various operating modes listed in (\ref{eqn:opt_sol}). This will yield insights on choice of $\alpha_w$ for buffer stability. We also describe the coin-toss probabilities and associated link-rates (buffered, direct and total) for various cases. In the second part of this section, we discuss performance with the two signalling schemes.
\subsubsection*{Average link-rate for a mode of operation}
\par Now, in order to obtain an expression  for average throughput, we need to represent the link rate of (\ref{eqn:taufin}) in a different form.  To this end, we first derive an expression for the joint probability $P_{R_{1}^{k_{1}}R_{2}^{k_{2}}R_{3}^{k_{3}}}$ 
of  $\{R_{1}^{k_{1}},R_{2}^{k_{2}},R_{3}^{k_{3}}\}\,(\equiv(k_{1},k_{2},k_{3}))$ being selected as the maximum permissible rates by links $1$, $2$ and $3$ respectively in any signalling interval. It is  defined as:
\begin{IEEEeqnarray}{rCl}\label{eqn:jointProb}
	\hspace{-.5cm}P_{R_{1}^{k_{1}}R_{2}^{k_{2}}R_{3}^{k_{3}}}&=&\Pr\{\underset{k_{\ell}}{\max}(I_{\U_{1}}^{k_{\ell}}R_{1}^{k_{\ell}})=R_{1}^{k_{1}}, \underset{k_{\ell}}{\max}(I_{\U_{2}}^{k_{\ell}}R_{2}^{k_{\ell}})=R_{2}^{k_{2}}, \underset{k_{\ell}}{\max}(I_{\U_{3}}^{k_{\ell}}R_{3}^{k_{\ell}})=R_{3}^{k_{3}}\}.
\end{IEEEeqnarray}
  It is clear from (\ref{eqn:opt_sol}) that link selection is associated with modes and range of indices $(k_1,k_2,k_3)$, we will find it convenient to study the 
 link-rate of link- $i$ associated with the mode $e$, which can be expressed as follows:
\begin{IEEEeqnarray}{rCl}\label{eqn:link_rate_domainset}
	\mcal{R}_{i}^{e}(\alpha_{w})=\underset{\mcal{U}^{e}(\alpha_{w})}{\sum}\hspace{-.2cm}P_{R_{1}^{k_{1}}R_{2}^{k_{2}}R_{3}^{k_{3}}}R_{i}^{k_{i}},
\end{IEEEeqnarray}
where the link-rate is averaged over the domain set for the given mode $e$ denoted by \sm{$\mcal{U}^{e}(\alpha_{w})$} (note that  \sm{$\mcal{U}^{e}(\alpha_{w})$} is the collection of all the index-triplets $(k_{1},k_{2},k_{3})$ associated with that mode). Hence we express the domain-set for all the possible modes as follows:
{\sm{
\begin{IEEEeqnarray}{rCl}\label{eqn:dom_set_discrete}
	\begin{array}{lll}
	\mcal{U}^{\N}(\alpha_{w}) \equiv \{(k_{1},k_{2},k_{3})|\, \alpha_{w} R_{1}^{k_{1}}= (1-\alpha_{w})R_{2}^{k_{2}}= R_{3}^{k_{3}}=0\},\\
	\mcal{U}^{i}(\alpha_{w}) \equiv \{(k_{1},k_{2},k_{3})|\, \varUpsilon_{i}^{k_i*}>\underset{j}{\max}(\varUpsilon_{j\neq i}^{k_j*})\},\ \ i\in \{1,2,3\}\\
	\mcal{U}^{ \widetilde{i}}(\alpha_{w}) \equiv \{(k_{1},k_{2},k_{3})|\, \varUpsilon_{j_{1}\neq i}^{k_{j_1}*}=\varUpsilon_{j_{2}\neq i}^{k_{j_2}*}>\varUpsilon_{i}^{k_i*}\}, i\in \{1,2,3\}\\
	\mcal{U}^{\stnn}(\alpha_{w}) \equiv \{(k_{1},k_{2},k_{3})|\, \alpha_{w} R_{1}^{k_{1}}= (1-\alpha_{w})R_{2}^{k_{2}}= R_{3}^{k_{3}}>0\}.
	\end{array}
\end{IEEEeqnarray}
} 
We will use the above to balance the buffer and to derive an expression for the  total system throughput.
\subsubsection*{Set of possible discrete $\alpha_{w}$ values}
\begin{figure}[!b]
	\begin{center}
		\includegraphics[width=160mm,height=60mm]{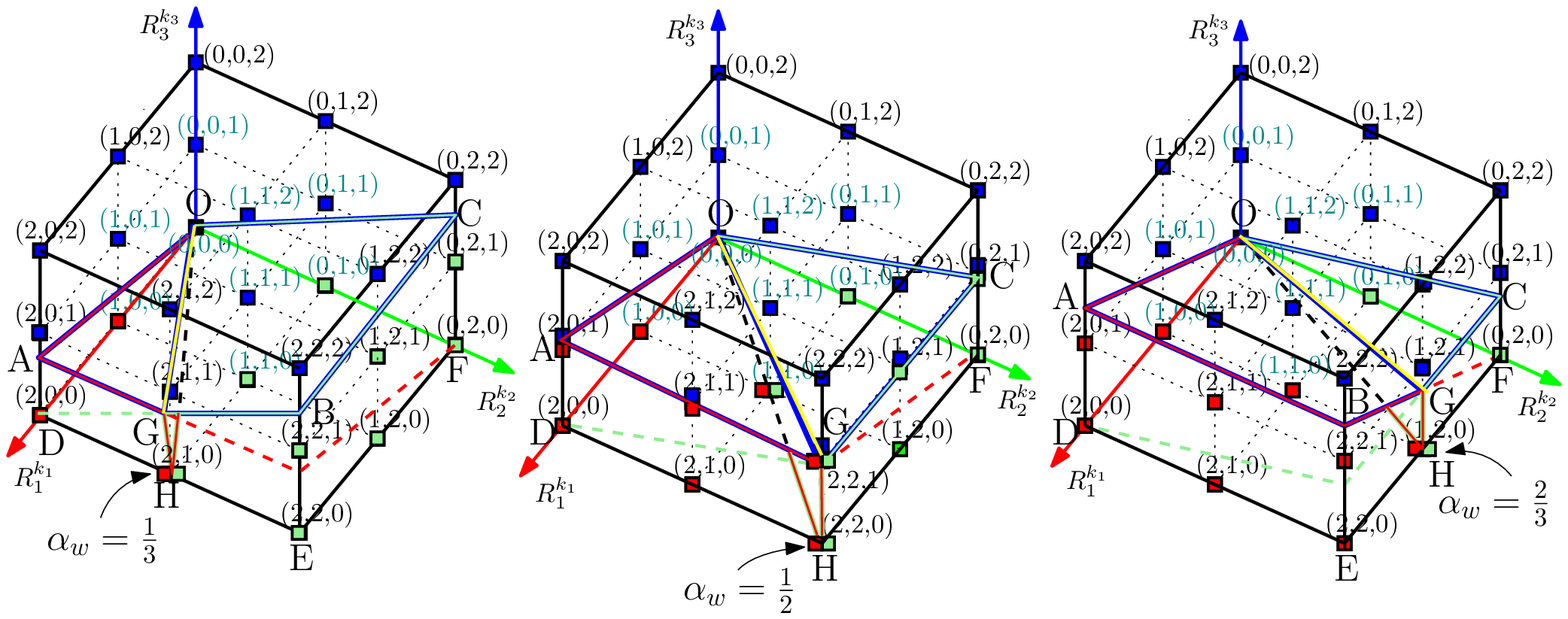}
		\caption{Rate triplets represented as a cubic lattice constellation  for $R_{1}^{k_{1}}=R_{2}^{k_{2}}=R_{3}^{k_{3}}=\{0,1,2\}$ in scheme-1. }
		\label{fig:constellation_cubic}
		\includegraphics[width=160mm,height=60mm]{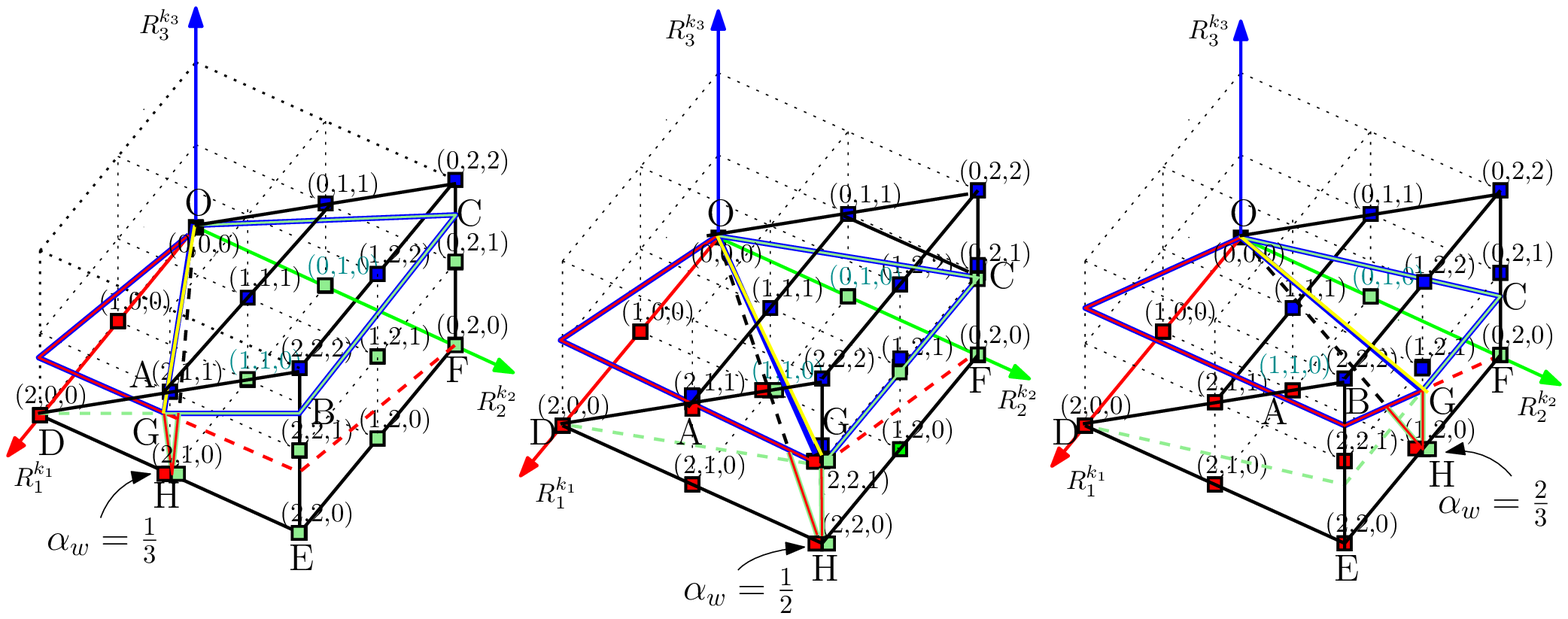}
		\caption{Rate triplets represented as a prismatic lattice constellation  for $R_{1}^{k_{1}}=R_{2}^{k_{2}}=R_{3}^{k_{3}}=\{0,1,2\}$  in scheme-2. }
		\label{fig:constellation_prism}
	\end{center}
	\vspace{-0.25cm}
\end{figure}
\par Since there are set of discrete rates available at $\SSS$ and $\SR$, it can be seen from (\ref{eqn:varUpsilon}) and (\ref{eqn:varUpsilon2})  that $\alpha_w$ takes discrete values.  We first observe all possible discrete values of $\alpha_{w}$ from domain set $\mcal{U}^{\stonen}(\alpha_{w}),\, \mcal{U}^{\stthrn}(\alpha_{w})$ and $\mcal{U}^{\sttwon}(\alpha_{w})$ as follows:
\sm{
	\begin{IEEEeqnarray*}{rCl}\label{eqn:cho_lambda2}
		\begin{array}{lll}
			
			R_{3}^{k_{3}}=(1-\alpha_{w})R_{2}^{k_{2}}&\Rightarrow&\alpha_{w}=1-R_{3}^{k_{3}}/R_{2}^{k_{2}}\quad\text{for } \mcal{U}^{\stonen}(\alpha_{w}),\\
			
			\alpha_{w} R_{1}^{k_{1}}=(1-\alpha_{w})R_{2}^{k_{2}}&\Rightarrow&\alpha_{w}=R_{2}^{k_{2}}/(R_{2}^{k_{2}}+R_{1}^{k_{1}})\quad\text{for } \mcal{U}^{\stthrn}(\alpha_{w}),\\
			
			R_{3}^{k_{3}}=\alpha_{w} R_{1}^{k_{1}}&\Rightarrow&\alpha_{w}=R_{3}^{k_{3}}/R_{1}^{k_{1}}\quad\text{for } \mcal{U}^{\sttwon}(\alpha_{w}).

		\end{array}
	\end{IEEEeqnarray*}
}
We denote by $\Lambda$ the set of all possible $\alpha_{w}$ values. In order to constraint $0\leq\alpha_{w}\leq 1$, we always choose link-$3$ whenever $R_{3}^{k_{3}} > R_{1}^{k_{1}}$ and $R_{3}^{k_{3}} > R_{2}^{k_{2}}$   ($R_{3}^{k_{3}}>\max(R_{1}^{k_{1}},R_{2}^{k_{2}})$). 
We define the set of $\alpha_{w}$, i.e., $\Lambda$ as:
\sm{
	\begin{IEEEeqnarray}{rCl}\label{eqn:cho_lambda}
			\alpha_{w}\in{\Lambda}&\equiv& \text{sort}\{1-R_{3}^{k_{3}}/R_{2}^{k_{2}},R_{2}^{k_{2}}/(R_{2}^{k_{2}}+R_{1}^{k_{1}}),R_{3}^{k_{3}}/R_{1}^{k_{1}}\hspace{0.5cm}:s.t.\, R_{3}^{k_{3}}\leq\max(R_{1}^{k_{1}},R_{2}^{k_{2}})\},\nonumber\\
			&\equiv&\{\alpha_{0}=0,\alpha_{1},\alpha_{2}....\alpha_{W-1},\alpha_{W}=1\}.
	\end{IEEEeqnarray}
}
where $\text{sort}\{x\}$ arranges  elements of set ${x}$ in increasing order.
\par {\bf\emph Example:} Consider an example with rate set $R_{1}^{[0,1,2]}=R_{2}^{[0,1,2]}=\{0,1,2\}$} (rate set  $R_{3}^{[0,1,2]}= R_{1}^{[0,1,2]}$ is implied). It is clear from (\ref{eqn:cho_lambda}) that  $\alpha_{w}\in {\Lambda}\equiv\{0,\frac{1}{3},\frac{1}{2},\frac{2}{3},1\}$. The rate triplets can be visualized as  points on a lattice constellation.  Fig. \ref{fig:constellation_cubic} and \ref{fig:constellation_prism} depict the rate triplets in the form of the constellation diagrams for scheme-1 and 2, which are in cube and prism shapes respectively. Furthermore, the regions belonging to three different links for $\alpha_{w}$ equal to $1/3$, $1/2$, and $2/3$ are depicted for both the schemes. Red, green and blue constellation points correspond to indices that lead to selection of link-$1$, $2$ and $3$ respectively. It is clear from the figures that the intersecting planes for  $\mcal{U}^{\stthrn}$ is always OGHO, and the point  $(2,2,1)$ lies on it for $\alpha_{w}=1/2$. Again for $\alpha_{w}=1/2$, the planes OAGO and OCGO belong to $\mcal{U}^{\sttwon}$ and $\mcal{U}^{\stonen}$ respectively. For  scheme-1, the planes OGHO, OAGO and OCGO generate the two rectangular pyramids for link-1 and link-2, whose bases are AGHDA and CGHFC. Any constellation point which lies inside any of the two pyramids belongs to the respective link. When we choose $\alpha_{w}=1/3$, the region belonging to link-$1$ is still a rectangular pyramid with base AGHDA, whereas the region belonging to link-$2$ changes to a trapezoidal pyramid, whose base is BCFEB. The region belonging to link-$3$ is the rest of the rectangular pyramid, generated by $\mcal{U}^{\stonen}$ and $\mcal{U}^{\sttwon}$. Similar arguments are valid for $\alpha_{w}=2/3$. For scheme-2, everything remains the same except that the rate triplets for which $k_{2}<k_{3}$ are no longer relevant.
\subsubsection*{Properties of link rate over domain set, i.e. $\mcal{R}_{i}^{e}(\alpha_{w})$}
\par We have already defined the link rate of link $i$, over the domain set $\sm{\mcal{U}^{e}(\alpha_{w})}$ in (\ref{eqn:link_rate_domainset}). The next lemma addresses two properties of these rates. The first property describes  relations between the rates in the events involving a coin-toss,  and the second investigates how rate continuity is maintained when $\alpha_{w}$ changes to $\alpha_{w+1}$ or $\alpha_{w-1}$. These properties will be used  later to determine the $\alpha_w$ that can stabilize the buffer, and to derive throughput expressions. \\
%
%
\par\emph{{\bf\emph Lemma 1:} The link rates over the domain set have two important properties as follows:\\ 
Property 1 (Rate-equality property): The link rates over the domain sets $\mcal{U}^{\stonen}(\alpha_{w}),\, \mcal{U}^{\sttwon}(\alpha_{w}),\,\mcal{U}^{\stthrn}(\alpha_{w})$ and $\mcal{U}^{\stnn}(\alpha_{w})$, which involve coin toss satisfy the following relations:
\begin{IEEEeqnarray}{rCl}\label{eqn:proj}
	\begin{array}{lll}
		\mcal{R}^{\stonen}_{3}(\alpha_{w})&=&(1-\alpha_{w})\mcal{R}^{\stonen}_{2}(\alpha_{w}),\, \mcal{R}^{\sttwon}_{3}(\alpha_{w})=\alpha_{w} \mcal{R}^{\sttwon}_{1}(\alpha_{w}),\\
		\alpha_{w} \mcal{R}^{\stthrn}_{1}(\alpha_{w})&=&(1-\alpha_{w})\mcal{R}^{\stthrn}_{2}(\alpha_{w}),\\
		\mcal{R}^{\stnn}_{3}(\alpha_{w})&=&\alpha_{w} \mcal{R}^{\stnn}_{1}(\alpha_{w})=(1-\alpha_{w})\mcal{R}^{\stnn}_{2}(\alpha_{w}).
	\end{array}		
\end{IEEEeqnarray}
{Property 2 (Rate-continuity property):} For $w\in\{1,2,...,W-1\}$, the following recursive relations hold:
\begin{IEEEeqnarray}{rCl}\label{eqn:linkrate_cont}
\begin{array}{lll}
	\mcal{R}^{\{1,\sttwon,\stthrn,\stnn\}}_{1}(\alpha_{w})=\mcal{R}_{1}^{1}(\alpha_{w+1}), \quad
	\mcal{R}^{\{2,\stonen,\stthrn,\stnn\}}_{2}(\alpha_{w})=  \mcal{R}_{2}^{2}(\alpha_{w-1}),\quad
	\mcal{R}^{\{3,\sttwon\}}_{3}(\alpha_{w})=\mcal{R}^{\{3,\stonen\}}_{3}(\alpha_{w-1}),
	\end{array}	
\end{IEEEeqnarray}
and we also have $\mcal{R}^{\{1,\sttwon,\stthrn,\stnn\}}_{1}(\alpha_{0}=0)=\mcal{R}^{\{2,\stonen,\stthrn,\stnn\}}_{2}(\alpha_{W}=1)=0.$\\
Proof: Property 1:} We prove $\mcal{R}^{\sttwon}_{3}(\alpha_{w})=\alpha_{w} \mcal{R}^{\sttwon}_{1}(\alpha_{w})$ first. We write $\mcal{R}^{\sttwon}_{3}(\alpha_{w})$ from (\ref{eqn:link_rate_domainset}) as follows:
\begin{IEEEeqnarray*}{rCl}
	\mcal{R}_{3}^{\sttwon}(\alpha_{w})&=&\underset{\mcal{U}^{\sttwon}(\alpha_{w})}{\sum}\hspace{-.2cm}P_{R_{1}^{k_{1}}R_{2}^{k_{2}}R_{3}^{k_{3}}}R_{3}^{k_{3}}=\underset{\mcal{U}^{\sttwon}(\alpha_{w})}{\sum}\hspace{-.2cm}P_{R_{1}^{k_{1}}R_{2}^{k_{2}}R_{3}^{k_{3}}}\alpha_{w}R_{1}^{k_{1}}
	=\alpha_{w}\underset{\mcal{U}^{\sttwon}(\alpha_{w})}{\sum}\hspace{-.2cm}P_{R_{1}^{k_{1}}R_{2}^{k_{2}}R_{3}^{k_{3}}}R_{1}^{k_{1}}=\alpha_{w} \mcal{R}^{\sttwon}_{1}(\alpha_{w}),
\end{IEEEeqnarray*}
where it is clear from (\ref{eqn:dom_set_discrete}) that the relation $R_{3}^{k_{3}}=\alpha_{w}R_{1}^{k_{1}}$ holds true for the domain set $\mcal{U}^{\sttwon}(\alpha_{w})$. The other relations can be proved in a similar manner.
\par \emph{Property 2}: We prove  $\mcal{R}^{\{1,\sttwon,\stthrn,\stnn\}}_{1}(\alpha_{w})=\mcal{R}_{1}^{1}(\alpha_{w+1})$ first.  It is clear from (\ref{eqn:dom_set_discrete}) that following relation holds true:
\begin{IEEEeqnarray*}{rCl}
	\begin{array}{lll}
		\mcal{U}^{ 1}(\alpha_{w}) \equiv \{(k_{1},k_{2},k_{3})|\,  \alpha_{w}R_{1}^{k_{1}}>\max(R_{3}^{k_{3}},(1-\alpha_{w})R_{2}^{k_{2}})\},\\
		\mcal{U}^{ \widetilde{2}}(\alpha_{w}) \equiv \{(k_{1},k_{2},k_{3})|\,  \alpha_{w}R_{1}^{k_{1}}=R_{3}^{k_{3}}> (1-\alpha_{w})R_{2}^{k_{2}}\},\\
		\mcal{U}^{ \widetilde{3}}(\alpha_{w}) \equiv \{(k_{1},k_{2},k_{3})|\,  \alpha_{w}R_{1}^{k_{1}}=(1-\alpha_{w})R_{2}^{k_{2}}>R_{3}^{k_{3}}\},\\
		\mcal{U}^{ \widetilde{N}}(\alpha_{w}) \equiv \{(k_{1},k_{2},k_{3})|\,  \alpha_{w}R_{1}^{k_{1}}=R_{3}^{k_{3}}= (1-\alpha_{w})R_{2}^{k_{2}}\}.
	\end{array}
\end{IEEEeqnarray*}
Since $\alpha_{w-1}<\alpha_{w}<\alpha_{w+1}$, it is apparent from the above equation that if we increase $\alpha_{w}$ to $\alpha_{w+1}$, all these above mentioned domain-sets transform into the following single domain-set:
\begin{IEEEeqnarray*}{rCl}
	\begin{array}{lll}
		\mcal{U}^{ 1}(\alpha_{w+1}) &\equiv& \{(k_{1},k_{2},k_{3})|\,  \alpha_{w+1}R_{1}^{k_{1}}> \max\left(R_{3}^{k_{3}}, (1-\alpha_{w+1})R_{2}^{k_{2}}\right)\}.
	\end{array}
\end{IEEEeqnarray*}
Hence, it is clear that $\mcal{U}^{\{1,\sttwon,\stthrn,\stnn\}}(\alpha_{w})$ and $\mcal{U}^{ 1}(\alpha_{w+1})$ are equal sets which we denote by 	$\mcal{U}^{1}(\alpha_{w+1})\equiv\mcal{U}^{\{1,\sttwon,\stthrn,\stnn\}}(\alpha_{w})$. 
It is also clear that if $\alpha_{0}=0$, then $\mcal{U}^{\{1,\sttwon,\stthrn,\stnn\}}(\alpha_{0}=0) \equiv \{\phi\}$ holds true, where $\phi$ is a null set. Hence, the following outcome is obvious:
{\normalsize
	\begin{IEEEeqnarray*}{rCl}
		\mcal{U}^{\{1,\sttwon,\stthrn,\stnn\}}(\alpha_{w})\equiv\mcal{U}^{1}(\alpha_{w+1})&\Rightarrow& \mcal{R}^{\{1,\sttwon,\stthrn,\stnn\}}_{1}(\alpha_{w})=\mcal{R}_{1}^{1}(\alpha_{w+1});\\
		\mcal{U}^{\{1,\sttwon,\stthrn,\stnn\}}(\alpha_{0}=0)\equiv \{\phi\}&\Rightarrow&  \mcal{R}^{\{1,\sttwon,\stthrn,\stnn\}}_{1}(\alpha_{0}=0)=0.
	\end{IEEEeqnarray*}
}\noindent
Similarly, other relations can be proved by the following inferences:
{\normalsize
	\begin{IEEEeqnarray*}{rCl}
		\mcal{U}^{\{2,\stonen,\stthrn,\stnn\}}(\alpha_{w})\equiv\mcal{U}^{2}(\alpha_{w-1})&\Rightarrow& \mcal{R}^{\{2,\stonen,\stthrn,\stnn\}}_{2}(\alpha_{w})=\mcal{R}_{2}^{2}(\alpha_{w-1});\\
		\mcal{U}^{\{2,\stonen,\stthrn,\stnn\}}(\alpha_{W}=1)\equiv \{\phi\}&\Rightarrow&  \mcal{R}^{\{2,\stonen,\stthrn,\stnn\}}_{2}(\alpha_{W}=1)=0;\\
		\mcal{U}^{\{3,\sttwon\}}(\alpha_{w})\equiv\mcal{U}^{\{3,\stonen\}}(\alpha_{w-1})&\Rightarrow&\mcal{R}^{\{3,\sttwon\}}_{3}(\alpha_{w})=\mcal{R}^{\{3,\stonen\}}_{3}(\alpha_{w-1}). w\neq0{\tiny\blacksquare}
	\end{IEEEeqnarray*}
}\noindent
{\bf\emph  Example (Contd.):} Consider the previous example for $\alpha_{w}=1/2$ in scheme-1 to understand these two properties. First, the following domain-sets are also evident from Fig. \ref{fig:constellation_cubic}:
\begin{IEEEeqnarray*}{rCl}
	\begin{array}{lll}
		\mcal{U}^{1}(1/2)\equiv\{(1,0,0),(2,0,0),\,(2,1,0)\},\ \mcal{U}^{2}(1/2)\equiv\{(0,1,0),(0,2,0),\,(1,2,0)\},\\
		\mcal{U}^{\stonen}(1/2)\equiv\{(0,2,1),\,(1,2,1)\},\,\mcal{U}^{\sttwon}(1/2)\equiv\{(2,0,1),\,(2,1,1)\},\,\mcal{U}^{\stthrn}(1/2)\equiv\{(1,1,0),\,(2,2,0)\},\\
		\mcal{U}^{\stnn}(1/2)\equiv\{(2,2,1)\}.
	\end{array}
\end{IEEEeqnarray*}
Suppose we want to validate $\alpha_{w}\mcal{R}^{\stthrn}_{1}(\alpha_{w})=(1-\alpha_{w})\mcal{R}^{\stthrn}_{2}(\alpha_{w})$ for $\alpha=1/2$ ($\mcal{R}^{\stthrn}_{1}(1/2)=\mcal{R}^{\stthrn}_{2}(1/2)$), which is based on the domain set $\mcal{U}^{\stthrn}(\alpha_{w}=1/2)\equiv\{(1,1,0),\,(2,2,0)\}$, the link-rates $\mcal{R}^{\stthrn}_{1}(1/2)$ and $\mcal{R}^{\stthrn}_{2}(1/2)$ are written as follows, and $\alpha_{w}\mcal{R}^{\stthrn}_{1}(\alpha_{w})=(1-\alpha_{w})\mcal{R}^{\stthrn}_{2}(\alpha_{w})$ holds for $\alpha=1/2$:
\begin{IEEEeqnarray*}{rCl}
	\begin{array}{lll}
		 \mcal{R}^{\stthrn}_{1}(1/2)=P_{R_{1}^{1}R_{2}^{1}R_{3}^{0}} R_{1}^{1}+P_{R_{1}^{2}R_{2}^{2}R_{3}^{0}}R_{1}^{2}=P_{1,1,0} +2P_{2,2,0},\\
		 \mcal{R}^{\stthrn}_{2}(1/2)=P_{R_{1}^{1}R_{2}^{1}R_{3}^{0}} R_{2}^{1}+P_{R_{1}^{2}R_{2}^{2}R_{3}^{0}}R_{2}^{2}=P_{1,1,0} +2P_{2,2,0}.
	\end{array}
\end{IEEEeqnarray*}
From property 2, it is inferred that  $\mcal{U}^{1}(2/3)\equiv\mcal{U}^{\{1,\sttwon,\stthrn,\stnn\}}(1/2)$ and $\mcal{U}^{2}(1/3)\equiv \mcal{U}^{\{2,\stonen,\stthrn,\stnn\}}(1/2)$ whereas $\mcal{U}^{\{1,\sttwon,\stthrn,\stnn\}}(1/2)$ and $\mcal{U}^{\{2,\stonen,\stthrn,\stnn\}}(1/2)$ can be written as:
\begin{IEEEeqnarray*}{rCl}
	\begin{array}{lll}
		\mcal{U}^{\{1,\sttwon,\stthrn,\stnn\}}(1/2)\equiv\{\overbrace{(1,0,0),(2,0,0),\,(2,1,0)}^{\mcal{U}^{1}(1/2)},\,\overbrace{(2,0,1),\,(2,1,1)}^{\mcal{U}^{\sttwon}(1/2)},\,\overbrace{(1,1,0),\,(2,2,0)}^{\mcal{U}^{\stthrn}(1/2)},\,\overbrace{(2,2,1)}^{\mcal{U}^{\stnn}(1/2)}\},\\
		\mcal{U}^{\{2,\stonen,\stthrn,\stnn\}}(1/2)\equiv\{\underbrace{(0,1,0),(0,2,0),\,(1,2,0)}_{\mcal{U}^{2}(1/2)},\,\underbrace{(0,2,1),\,(1,2,1)}_{\mcal{U}^{\stonen}(1/2)},\,\underbrace{(1,1,0),\,(2,2,0)}_{\mcal{U}^{\stthrn}(1/2)},\,\underbrace{(2,2,1)}_{\mcal{U}^{\stnn}(1/2)}\}.
	\end{array}
\end{IEEEeqnarray*}
However, $(2,0,1)$ (for which $k_{2}<k_{3}$) is invalid rate triplet for scheme-2 in this example.
\subsubsection*{Average link-rate for link-i}
\par We now express the average link-rate in terms of link-rates of possible modes for link-$i$. It is evident from (\ref{eqn:CT_Prob1}) that there are  several coin toss probabilities. It will be shown later in  lemma-3 of this section that the overall throughput does not depend on these coin toss probabilities. Instead, these probabilities only alter individual link rates.
We can associate the coin toss probabilities of mode $\stnn$ with modes $\stonen$, $\sttwon$ or with $\stthrn$. 
Since associating these probabilities with mode $\stthrn$ might increase the buffer-usage in some cases, we associate them with modes $\stonen$ and $\sttwon$ as follows:
\begin{IEEEeqnarray*}{rCl}
	\begin{array}{lll}
		P^{\stnn}_{1}(\alpha_{w})=P^{\sttwon}_{1}(\alpha_{w})\quad \text{and}\quad  P^{\stnn}_{2}(\alpha_{w})=P^{\stonen}_{2}(\alpha_{w}).
	\end{array}
\end{IEEEeqnarray*}
The link rate of link-$i$, i.e. $\tau_{i}$ for $i\in\{1,2,3\}$, are now expressed as follows:
\begin{IEEEeqnarray}{rCl}\label{eqn:S_fun}
	&&\tau_{1}(\alpha_{w},P_1^{1},P^{\sttwon}_{1},P^{\stthrn}_{1})= P^{1}_{1}(\alpha_{w})\mcal{R}_{1}^{1}(\alpha_{w})+ P^{\sttwon}_{1}(\alpha_{w}) \mcal{R}^{\{\sttwon,\stnn\}}_{1}(\alpha_{w})+ P^{\stthrn}_{1}(\alpha_{w})\mcal{R}^{\stthrn}_{1}(\alpha_{w}),\nonumber\\
	&&\tau_{2}(\alpha_{w},P_2^{2},P^{\stonen}_{2},P^{\stthrn}_{2})= P^{2}_{2}(\alpha_{w})\mcal{R}_{2}^{2}(\alpha_{w})+ P^{\stonen}_{2}(\alpha_{w}) \mcal{R}^{\{\stonen,\stnn\}}_{2}(\alpha_{w})+ P^{\stthrn}_{2}(\alpha_{w})\mcal{R}^{\stthrn}_{2}(\alpha_{w}),\\
	&&\tau_{3}(\alpha_{w},P^{\sttwon}_{1},P^{\stonen}_{2})\hspace{-0.1cm}=\hspace{-0.1cm}\mcal{R}_{3}^{3}(\alpha_{w})+ \ovl{P}^{\sttwon}_{1}(\alpha_{w})\mcal{R}^{\sttwon}_{3}(\alpha_{w})+\ovl{P}^{\stonen}_{2}(\alpha_{w})\mcal{R}^{\stonen}_{3}(\alpha_{w})+\left(\ovl{P^{\stonen}_{2}(\alpha_{w})+P^{\sttwon}_{1}(\alpha_{w})}\right)\mcal{R}^{\stnn}_{3}(\alpha_{w}),\nonumber
\end{IEEEeqnarray}
where probabilities $P^{1}_{1}(\alpha_{w})$ and $P^{2}_{2}(\alpha_{w})$ are useful in some special scenarios described later in the paper. The above equation will be utilized (in what follows) to find an expression for the optimum system throughput and to establish the buffer-stability conditions in various scenarios.
%
\subsubsection*{Use cases for buffer stability}
In underlay cognitive radio networks, the  average SNR of a link is dependent on the  forward and the interference links to the primary receiver. For this reason, asymmetry in average SNRs of links is common in a two-hop network, even when the relay is located mid-way between the source and the destination. Three use cases are clearly of interest. In case-$1$, link-$2$ is heavily attenuated, whereas in case-$2$, link-$1$ is heavily attenuated. In case-$3$, neither link-$1$ nor link-$2$ is heavily attenuated. We discuss buffer-balancing conditions in these use cases.
\par \emph{{\bf\emph Lemma 2:}  The buffer is stable for $\alpha_{w}\leq \alpha_{w^{*}}$, with:
\begin{IEEEeqnarray}{rCl}\label{eqn:buf_conds}
	\begin{array}{lll}
		{w}^{*}\hspace{-0.25cm} &=&\hspace{-0.25cm}\left\{
		\begin{array}{ll}
			0 \hspace{1.4cm}  \mathrm{if }\  \mcal{R}_{2}^{2}(\alpha_{0})<\mcal{R}_{1}^{1}(\alpha_{1})\hspace{0.65cm}
			 \mathrm{:case\mhyphen 1}\\
			W-1 \quad \mathrm{if\  } \mcal{R}_{1}^{1}(\alpha_{W})<\mcal{R}_{2}^{2}(\alpha_{W-1})\mathrm{:case\mhyphen  2}\\
			z \hspace{1.4cm} \mathrm{if\  } \mcal{R}_{2}^{2}(\alpha_{z-1})\geq \mcal{R}_{1}^{1}(\alpha_{z})
			\hspace{.15cm}\&\, \mcal{R}_{1}^{1}(\alpha_{z+1})\geq \mcal{R}_{2}^{2}(\alpha_{z})\hspace{0.25cm}\mathrm{:case\mhyphen 3}
		\end{array}
		\right.
	\end{array}
\end{IEEEeqnarray}
where $z$ is an integer such that $z\in\{1,2,....,W-1\}$.\\
Proof:} We first consider case-$1$ and case-$2$. Using the rate continuity property $\mcal{R}^{\{2,\stonen,\stthrn,\stnn\}}_{2}(\alpha_{w})=  \mcal{R}_{2}^{2}(\alpha_{w-1})$ of (\ref{eqn:linkrate_cont}) with $w=1$, we infer the following: 
\begin{IEEEeqnarray*}{rCl}
	\begin{array}{lll}
		\mcal{R}^{\{2,\stonen,\stthrn,\stnn\}}_{2}(\alpha_{1})&=&  \mcal{R}_{2}^{2}(\alpha_{0})\,\,\Rightarrow \,\,
		\mcal{R}^{2}_{2}(\alpha_{1}) \leq \mcal{R}_{2}^{2}(\alpha_{0}).
	\end{array}	
\end{IEEEeqnarray*}
If we assume that link-2 is attenuated such that $\mcal{R}_{2}^{2}(\alpha_{1})<\mcal{R}_{1}^{\{1,\sttwon,\stthrn,\stnn\}}(\alpha_{1})$, then we assign all the coin-toss events to link-2, so that $\tau_{2}(\alpha_{1},1,0,0)=\mcal{R}_{2}^{2}(\alpha_{1})$ increases to $\tau_{2}(\alpha_{1},1,1,1)=\mcal{R}_{2}^{\{2,\stonen,\stthrn,\stnn\}}(\alpha_{1})$, which is equal to $R_{2}^{2}(\alpha_{0})$. We summarize this as follows:
\begin{IEEEeqnarray}{rCl}\label{eqn:state1}
	\begin{array}{lll}
		P^{\sttwon}_{1}(\alpha_{1})=\ovl{P}^{\stonen}_{2}(\alpha_{1})=P^{\stthrn}_{1}(\alpha_{1})=0\Rightarrow\\ \tau_{1}(\alpha_{1},1,0,0)=\mcal{R}_{1}^{1}(\alpha_{1}),\,\,\tau_{2}(\alpha_{1},1,1,1)=\mcal{R}_{2}^{\{2,\stonen,\stthrn,\stnn\}}(\alpha_{1})={\cal R}_{2}^{2}(\alpha_{0}),\,\,\tau_{3}(\alpha_{0},0,1)=\mcal{R}_{3}^{\{3,\sttwon\}}(\alpha_{1}).
	\end{array}	
\end{IEEEeqnarray}
If the link-2 is so heavily attenuated that condition  $\mcal{R}_{2}^{2}(\alpha_{0})<\mcal{R}_{1}^{1}(\alpha_{1})$ still holds, lowering $\alpha_w$ further in an attempt to stabilize the buffer is not feasible since there is no inflow rate. We summarize this state as follows:
\begin{IEEEeqnarray}{rCl}\label{eqn:state0}
	\begin{array}{lll}
		P^{\sttwon}_{1}(\alpha_{0})=\ovl{P}^{\stonen}_{2}(\alpha_{0})=P^{\stthrn}_{1}(\alpha_{0})=1\Rightarrow\\ \tau_{1}(\alpha_{0},1,1,1)=\mcal{R}_{1}^{\{1,\sttwon,\stthrn,\stnn\}}(\alpha_{0})=0,\,\,\tau_{2}(\alpha_{0},1,0,0)=\mcal{R}_{2}^{2}(\alpha_{0}),\,\,\tau_{3}(\alpha_{0},1,0)=\mcal{R}_{3}^{\{3,\stonen\}}(\alpha_{0}).
	\end{array}	
\end{IEEEeqnarray}
However, it is evident from (\ref{eqn:state1}) that we can change the link selection probability $P_{1}^{1}(\alpha_{1})$ to balance the buffer. Clearly, the buffered and direct throughput of $R_{2}^{2}(\alpha_{0})$ and $\mcal{R}_{3}^{\{3,\stonen\}}(\alpha_{0})$ can maximally be achieved and buffer can be balanced with $\alpha_{1}$ when $P_{1}^{1}(\alpha_{1})=\mcal{R}_{2}^{2}(\alpha_{0})/\mcal{R}_{1}^{1}(\alpha_{1})$. Similar arguments can be given for condition $\mcal{R}_{2}^{2}(\alpha_{w-1})< \mcal{R}_{1}^{1}(\alpha_{w})$ in case-2, when link-1 is heavily attenuated.  In this case,  the buffer throughput of $\mcal{R}_{1}^{1}(\alpha_{W})$ can maximally be achieved with direct throughput $\mcal{R}^{\{3,\stonen\}}_{3}(\alpha_{W-1})=\mcal{R}^{\{3,\sttwon\}}_{3}(\alpha_{W})$, and  buffer can be balanced with $\alpha_{W-1}$ when $P_{2}^{2}(\alpha_{W-1})=\mcal{R}_{1}^{1}(\alpha_{W})/\mcal{R}_{2}^{2}(\alpha_{W-1})$.
\par Now consider case-3, in which neither link-1 nor link-2 is heavily attenuated. It is apparent from (\ref{eqn:S_fun}) that for some $\alpha_{w^{*}}=\alpha_{z}$, if the conditions $\tau_{1}(\alpha_{z},1,1,1)\geq \tau_{2}(\alpha_{z},1,0,0)$ and $\tau_{1}(\alpha_{z},1,0,0)\leq \tau_{2}(\alpha_{z},1,1,1)$ are satisfied, the inflow rate is equal to that of the outflow for some combination of these coin toss probabilities.
After applying the rate-continuation property for link-1 and 2  for $z\in\{1,2,...,W-1\}$, we get the following:
\begin{IEEEeqnarray*}{rCl}
	\begin{array}{lll}
		\tau_{1}(\alpha_{z},1,1,1)=\mcal{R}^{\{1,\sttwon,\stthrn,\stnn\}}_{1}(\alpha_{z})=\mcal{R}_{1}^{1}(\alpha_{z+1}),\quad \tau_{1}(\alpha_{z},1,0,0)=\mcal{R}_{1}^{1}(\alpha_{z}),\\
		\tau_{2}(\alpha_{z},1,1,1)=\mcal{R}^{\{2,\stonen,\stthrn,\stnn\}}_{2}(\alpha_{z})=  \mcal{R}_{2}^{2}(\alpha_{z-1}),\quad \tau_{2}(\alpha_{z},1,0,0)=\mcal{R}_{2}^{2}(\alpha_{z}).
	\end{array}
\end{IEEEeqnarray*}
The following conditions hold from the above equation and the argument stated previously:
\begin{IEEEeqnarray}{rCl}\label{eqn:recur1}
	\begin{array}{lll}
		\mcal{R}_{2}^{2}(\alpha_{z-1})\geq \mcal{R}_{1}^{1}(\alpha_{z}) \quad\text{and}\quad \mcal{R}_{1}^{1}(\alpha_{z+1})\geq \mcal{R}_{2}^{2}(\alpha_{z}),
	\end{array}
\end{IEEEeqnarray}
which is given by (\ref{eqn:buf_conds}). Now we establish two recursions:
\begin{IEEEeqnarray*}{rCl}
		\begin{array}{lll}
			\mcal{R}_{2}^{2}(\alpha_{z-1})\geq \mcal{R}_{1}^{1}(\alpha_{z})
			\overset{l}{\Leftrightarrow}\,\,\mcal{R}_{2}^{2}(\alpha_{z-1})\geq \mcal{R}_{1}^{\{1,\sttwon,\stthrn,\stnn\}}(\alpha_{z-1})\\
			\overset{m}{\Rightarrow}\,\, \mcal{R}_{2}^{\{2,\stonen,\stthrn,\stnn\}}(\alpha_{z-1})\geq \mcal{R}_{1}^{1}(\alpha_{z-1})\overset{l}{\Leftrightarrow}\,\,\mcal{R}_{2}^{2}(\alpha_{z-2})\geq \mcal{R}_{1}^{1}(\alpha_{z-1}),
		\end{array}\\
		\begin{array}{lll}
			\mcal{R}_{1}^{1}(\alpha_{z+1})\geq \mcal{R}_{2}^{2}(\alpha_{z})
			\overset{l}{\Leftrightarrow}\,\, \mcal{R}_{1}^{1}(\alpha_{z+1})\geq \mcal{R}_{2}^{\{2,\stonen,\stthrn,\N\}}(\alpha_{z+1})\\
			\overset{m}{\Rightarrow}\,\, \mcal{R}_{1}^{\{1,\sttwon,\stthrn,\N\}}(\alpha_{z+1})\geq \mcal{R}_{2}^{2}(\alpha_{z+1})\overset{l}{\Leftrightarrow}\,\,\mcal{R}_{1}^{1}(\alpha_{z+2})\geq \mcal{R}_{2}^{2}(\alpha_{z+1}),
		\end{array}
\end{IEEEeqnarray*}
where $l$ is implied by the rate-continuity property and $m$ due to change in coin-toss probabilities $P^{\sttwon}_{1}(\alpha_{z})$, $\ovl{P}^{\stonen}_{2}(\alpha_{z}),\,P^{\stthrn}_{1}(\alpha_{z})$ from 1 to 0 or 0 to 1. It is clear from the above recursions that $	\mcal{R}_{2}^{2}(\alpha_{z-2})\geq \mcal{R}_{1}^{1}(\alpha_{z-1})$ and $ \mcal{R}_{1}^{1}(\alpha_{z+2})\geq \mcal{R}_{2}^{2}(\alpha_{z+1})$.
Hence, it is evident that the following will clearly hold true:
{\normalsize
	\begin{IEEEeqnarray}{rCl}\label{eqn:recur3}
		\begin{array}{lll}
			\mcal{R}_{2}^{2}(\alpha_{0})\geq \mcal{R}_{1}^{1}(\alpha_{1}) \quad\text{and}\quad \mcal{R}_{1}^{1}(\alpha_{W})\geq \mcal{R}_{2}^{2}(\alpha_{W-1}),
		\end{array}
	\end{IEEEeqnarray}
}\noindent
Note that the above contradict conditions for case-1 and case-2. This shows that when case-3 holds, case-1 and 2 can be ruled out. 
Now, in order to show that  $\alpha_{z}$ is unique, we re-write the conditions for $\alpha_{z-1}$, $\alpha_{z}$, and $\alpha_{z+1}$ together as follows: 
\begin{IEEEeqnarray}{rCl}\label{eqn:recur4}
	\begin{array}{lll}
		\mcal{R}_{2}^{2}(\alpha_{z-2})\geq \mcal{R}_{1}^{1}(\alpha_{z-1}) \quad\text{and}\quad \mcal{R}_{1}^{1}(\alpha_{z})\geq \mcal{R}_{2}^{2}(\alpha_{z-1}), \text{ for } \alpha_{z-1},\\
		\mcal{R}_{2}^{2}(\alpha_{z-1})\geq \mcal{R}_{1}^{1}(\alpha_{z}) \quad\text{and}\quad \mcal{R}_{1}^{1}(\alpha_{z+1})\geq \mcal{R}_{2}^{2}(\alpha_{z}), \quad\text{ for } \alpha_{z},\\
		\mcal{R}_{2}^{2}(\alpha_{z})\geq \mcal{R}_{1}^{1}(\alpha_{z+1}) \quad\text{and}\quad \mcal{R}_{1}^{1}(\alpha_{z+2})\geq \mcal{R}_{2}^{2}(\alpha_{z+1}), \text{ for } \alpha_{z+1}.		
	\end{array}
\end{IEEEeqnarray}
It is clear from the above that when the middle equation for $\alpha_{z}$ holds, the other two cannot hold simultaneously (the conditions are contradictory). Following these arguments, it can therefore be inferred that case-3 condition is indeed satisfied by a unique $\alpha_{z}$ only. $\blacksquare$
\subsubsection*{Expression for optimum system throughput $\tau_{t}$}
\par Now, in the next lemma we present an expression for optimum system throughput for the given system model, which is valid for both the used schemes.
\par\emph{{\bf\emph Lemma 3:}  The average throughput of the system can be written in a simplified form as:
\begin{IEEEeqnarray}{rCl}\label{eqn:sys_ratee}
	\begin{array}{lll}
	\hspace{-.0cm}\tau_{t}&=&\underset{\alpha_{w}\in \Lambda}{\min}\Big(\alpha_{w}\mcal{R}^{\{1,\sttwon,\stnn,\stthrn\}}_{1}(\alpha_{w})+(1-\alpha_{w})\mcal{R}_{2}^{2}(\alpha_{w})+ \mcal{R}^{\{3,\stonen\}}_{3}(\alpha_{w})\Big), \quad w\neq W\\ 
	\hspace{-.0cm} &=&\underset{\alpha_{w}\in \Lambda}{\min}\Big(\alpha_{w}\mcal{R}_{1}^{1}(\alpha_{w})+(1-\alpha_{w})\mcal{R}^{\{2,\stonen,\stnn,\stthrn\}}_{2}(\alpha_{w})+\mcal{R}^{\{3,\sttwon\}}_{3}(\alpha_{w})\Big). \quad w\neq 0 
	\end{array}
\end{IEEEeqnarray}
Proof:} Please see Appendix A.$\blacksquare$
\par\emph{{\bf\emph Remark 1:} It is clear from (\ref{eqn:sys_ratee}) that in calculating the optimum average rate of the system, the inflow, outflow and direct link rates are weighed by $\alpha_{w^{*}},\,1-\alpha_{w^{*}}$ and $1$ due to buffer balancing. 
The optimum value of $\alpha_w$ was deterxmined using lemma-2. It should be noted that (\ref{eqn:sys_ratee}) can also be used to determine the optimum value of $\alpha_{w}$ by looking for the value of $\alpha_{w}$ for which the terms in the brackets of right-hand side is minimized.}
\par\emph{{\bf\emph Remark 2:} There are many combinations of coin-toss probabilities which leads to the same optimum solution of system throughput with a balanced buffer. The optimum throughput of the balanced buffer and the direct path might change, but optimum system throughput remains the same for these coin-toss probability combinations. It is apparent from (\ref{eqn:lagrangian}) that coin-toss probabilities assist in balancing the buffer, not in maximizing throughput of the system.}\\ 
we now discuss the buffered/direct throughput and coin-toss probabilities next.
\subsubsection*{Expression of coin-toss probabilities and link throughput}
There exist more than one unique combination of coin-toss probabilities that yield the same optimum system throughput. For case-1 and 2, we have already discussed about the choice of $P^{1}_{1}(\alpha_{1})$ and $P^{2}_{2}(\alpha_{W-1})$ and the relevant buffered/direct-link throughput. We now provide some analytical expressions for coin-toss probabilities in case-3. The buffer is balanced by suitable choice of $P^{\stthrn}_{1}(\alpha_{z})$  (and thereby $P^{\stthrn}_{2}(\alpha_{z})=1-P^{\stthrn}_{1}(\alpha_{z})$) when either link-1 or link-2 are relatively weak (while not being weak enough to belong to case-1 or case-2 ). Three subcases arise as listed in Table II. In case-3a, link-2 is relatively weak so that ${\cal R}_1^1(\alpha_z)\geq {\cal R}_2^{\{\widetilde{2},\widetilde{3}\}}(\alpha_z)$ and ${\cal R}_1^{\{1,\widetilde{3}\}}(\alpha_z)\geq {\cal R}_2^2(\alpha_z)$.  In this case, we set  both $P^{\sttwon}_{1}(\alpha_{z})$ and  $P^{\stonen}_{2}(\alpha_{z})$  to zero and use $P^{\stthrn}_{1}(\alpha_{z})$ (and thereby $P^{\stthrn}_{2}(\alpha_{z})$) to balance the buffer (note that this might reduce throughput of the direct path). The choice of $P_1^{\widetilde{2}}(\alpha_z)$ then follows from (\ref{eqn:S_fun}). Case-3c follows similarly when link-1 is relatively weak. Case-3b arises when ${\cal R}_2^{\{2,\widetilde{3}\}}(\alpha_z)\geq {\cal R}_1^{1}(\alpha_z)$ and ${\cal R}_1^{\{1,\widetilde{3}\}}(\alpha_z)\geq {\cal R}_2^{2}(\alpha_z)$. The fourth sub-case does not exist because ${\mcal{R}_{1}^{1}(\alpha_{w})\geq \mcal{R}^{\{2,\stthrn\}}_{2}(\alpha_{w})}$ implies ${\mcal{R}_{1}^{\{1,\stthrn\}}(\alpha_{w})\geq \mcal{R}^{2}_{2}(\alpha_{w})}$ and vice-versa. 

\begin{table*}[!h]
\caption{Different use-cases, their conditions, coin-toss probabilities and buffered/direct throughputs}
\centering
\hspace{1cm}
\renewcommand{\arraystretch}{1.2}
\label{tab:usecasecointosstau} 
\begin{minipage}{\textwidth}
	\resizebox{\textwidth}{!}{\begin{tabular}{|p{.5cm}|p{5.75 cm}|p{4.75cm}|p{9cm}|}
	\hline
	Use case & \hspace{1.75cm}Condition &   \hspace{.5cm}Coin-toss Probability  & \hspace{.5cm}(Buffered/Direct) throughput (using (\ref{eqn:S_fun}))\\\hline
	\multirow {2}{*}{1} 
	&\vspace{-.25cm} \hspace{-.15cm}$\mcal{R}_{2}^{2}(\alpha_{0})<\mcal{R}_{1}^{1}(\alpha_{1})$   
	&\vspace{-.25cm} $(\alpha_{0}):P^{\sttwon}_{1}=\ovl{P}^{\stonen}_{2}=P^{\stthrn}_{1}=1$ 
	&\vspace{-.25cm} \hspace{-.15cm}$\tau_{1}=\tau_{2}=\mcal{R}_{2}^{\{2,\stonen,\stthrn,\stnn\}}(\alpha_{1})=\mcal{R}_{2}^{2}(\alpha_{0})$  \\
				
	&  \hspace{-.15cm}with $P_{1}^{1}(\alpha_{1})=\frac{\mcal{R}_{2}^{2}(\alpha_{0})}{\mcal{R}_{1}^{1}(\alpha_{1})}$	 	
	&  $(\alpha_{1}): P^{\sttwon}_{1}=\ovl{P}^{\stonen}_{2}=P^{\stthrn}_{1}=0$			
	&  \hspace{-.15cm}$\tau_{3}=\mcal{R}^{\{3,\sttwon\}}_{3}(\alpha_{1})=\mcal{R}^{\{3,\stonen\}}_{3}(\alpha_{0})$\\\hline
		
	\multirow {2}{*}{2}	&\vspace{-.25cm}\hspace{-.15cm} $\mcal{R}_{1}^{1}(\alpha_{W})<\mcal{R}_{2}^{2}(\alpha_{W-1})$	
	&\vspace{-.25cm}\hspace{-.25cm} $ (\alpha_{W-1})\hspace{-.08cm}:\hspace{-.1cm} P^{\sttwon}_{1}=\ovl{P}^{\stonen}_{2}=P^{\stthrn}_{1}=1$ 
	&\vspace{-.25cm} \hspace{-.15cm}$\tau_{1}=\tau_{2}=\mcal{R}_{1}^{\{1,\sttwon,\stthrn,\stnn\}}(\alpha_{W-1})=\mcal{R}_{1}^{1}(\alpha_{W})$\\
		
	&\hspace{-.15cm} with  $P_{2}^{2}(\alpha_{W-1})=\frac{\mcal{R}_{1}^{1}(\alpha_{W})}{\mcal{R}_{2}^{2}(\alpha_{W-1})}$	
	& $ (\alpha_{W}): P^{\sttwon}_{1}=\ovl{P}^{\stonen}_{2}=P^{\stthrn}_{1}=0$			
	&  \hspace{-.15cm}$\tau_{3}=\mcal{R}^{\{3,\stonen\}}_{3}(\alpha_{W-1})=\mcal{R}^{\{3,\sttwon\}}_{3}(\alpha_{W})$\\\hline
			
	\multirow {2}{*}{3a} &\vspace{-.25cm}\hspace{-.25cm} $\mcal{R}^{2}_{2}(\alpha_{z-1})\geq {\mcal{R}_{1}^{1}(\alpha_{z})\geq \mcal{R}^{\{2,\stthrn\}}_{2}(\alpha_{z})}$ 
	& \vspace{-.25cm} $(\alpha_{z}):P^{\sttwon}_{1}=P^{\stthrn}_{1}=0$
	&\vspace{-.25cm} \hspace{-.15cm}$\tau_{1}=\tau_{2}=\mcal{R}_{1}^{1}(\alpha_{z})$\\
				
	& \hspace{-.15cm}$\mcal{R}^{1}_{1}(\alpha_{z+1})\geq \mcal{R}^{\{1,\stthrn\}}_{1}(\alpha_{z})\geq\mcal{R}_{2}^{2}(\alpha_{z}) $		 	
	& $P^{\stonen}_{2}=\frac{\mcal{R}_{1}^{1}(\alpha_{z})-\mcal{R}^{\{2,\stthrn\}}_{2}(\alpha_{z})}{\mcal{R}^{\{\stonen,\stnn\}}_{2}(\alpha_{z})}$
	& \hspace{-.15cm}$\tau_{3}=\mcal{R}^{\{3,\stonen,\sttwon,\stnn\}}_{3}(\alpha_{z})-(1-\alpha_{z})(\mcal{R}_{1}^{1}(\alpha_{z})-\mcal{R}^{\{2,\stthrn\}}_{2}(\alpha_{z}))$\\\hline
		
	\multirow {3}{*}{3b}	&\vspace{-.25cm}\hspace{-.25cm} $\mcal{R}^{2}_{2}(\alpha_{z-1})\geq \mcal{R}^{\{2,\stthrn\}}_{2}(\alpha_{z})\geq \mcal{R}_{1}^{1}(\alpha_{z})$ 
	&\vspace{-.25cm}  $(\alpha_{z}):P^{\sttwon}_{1}=P^{\stonen}_{2}=0$
	& \vspace{-.25cm}\hspace{-.15cm}$\tau_{1}=\tau_{2}=\mcal{R}_{1}^{1}(\alpha_{z})+(1-\alpha_{z})(\mcal{R}^{\{2,\stthrn\}}_{2}(\alpha_{z})-\mcal{R}_{1}^{1}(\alpha_{z}))$ \\
	&&   $P^{\stthrn}_{1}=\frac{\mcal{R}^{\{2,\stthrn\}}_{2}(\alpha_{w})-\mcal{R}_{1}^{1}(\alpha_{z})}{\mcal{R}^{\stthrn}_{1}(\alpha_{z})+\mcal{R}^{\stthrn}_{2}(\alpha_{z})}$
	&  \hspace{-.15cm}$\tau_{1}=\tau_{2}=\mcal{R}_{2}^{2}(\alpha_{z})+\alpha_{z}(\mcal{R}^{\{1,\stthrn\}}_{1}(\alpha_{z})-\mcal{R}_{2}^{2}(\alpha_{z}))$ \\
	& \hspace{-.15cm}$\mcal{R}^{1}_{1}(\alpha_{z+1})\geq \mcal{R}^{\{1,\stthrn\}}_{1}(\alpha_{z})\geq\mcal{R}_{2}^{2}(\alpha_{z}) $		 	
	&   $P^{\stthrn}_{2}=\frac{\mcal{R}^{\{1,\stthrn\}}_{1}(\alpha_{w})-\mcal{R}_{2}^{2}(\alpha_{z})}{\mcal{R}^{\stthrn}_{1}(\alpha_{z})+\mcal{R}^{\stthrn}_{2}(\alpha_{z})}$
	&\hspace{-.25cm} $\tau_{3}=\mcal{R}^{\{3,\stonen,\sttwon,\stnn\}}_{3}(\alpha_{z})$\\\hline
				
	\multirow {2}{*}{3c}	& \vspace{-.25cm} \hspace{-.15cm}$\mcal{R}^{2}_{2}(\alpha_{z-1})\geq \mcal{R}^{\{2,\stthrn\}}_{2}(\alpha_{z})\geq \mcal{R}_{1}^{1}(\alpha_{z})$ 
	&\vspace{-.25cm}  $(\alpha_{z}):P^{\stonen}_{2}=P^{\stthrn}_{2}=0$
	&\vspace{-.25cm}\hspace{-.15cm} $\tau_{1}=\tau_{2}=\mcal{R}_{2}^{2}(\alpha_{z})$\\
	& \hspace{-.15cm}$\mcal{R}^{1}_{1}(\alpha_{z+1})\geq\mcal{R}_{2}^{2}(\alpha_{z})\geq \mcal{R}^{\{1,\stthrn\}}_{1}(\alpha_{z}) $		 	
	&  $P^{\sttwon}_{1}=\frac{\mcal{R}_{2}^{2}(\alpha_{z})- \mcal{R}^{\{1,\stthrn\}}_{1}(\alpha_{z})}{\mcal{R}^{\{\sttwon,\stnn\}}_{1}(\alpha_{z})}$
	&  \hspace{-.15cm}$\tau_{3}=\mcal{R}^{\{3,\stonen,\sttwon,\stnn\}}_{3}(\alpha_{z})-\alpha_{z}(\mcal{R}_{2}^{2}(\alpha_{z})- \mcal{R}^{\{1,\stthrn\}}_{1}(\alpha_{z}))$\\\hline
\end{tabular}}
\vspace{-0.175cm}
\label{tab:CCDF_LSP} 
\end{minipage}
\end{table*}
\par\emph{\noindent {\bf\emph Lemma 4:} The relevant conditions for all possible cases that are formulated from the condition of buffer stability in (\ref{eqn:buf_conds}) are presented in Table-\ref{tab:usecasecointosstau}. The coin-toss probability and throughput of relevant cases, which are subsequently derived from (\ref{eqn:S_fun}), are also summarized in Table-\ref{tab:usecasecointosstau}.\\
Proof:} Use cases 1 and 2 have been discussed already.
The conditions for cases 3a, 3b, and 3c are mentioned in  Table-\ref{tab:usecasecointosstau}. We first prove case 3a,  and cases 3b and 3c can be proved in a similar fashion. In this subcase of case 3,  link-2 is weak enough  so that even after setting $P^{\sttwon}_{1}(\alpha_{z})=P^{\stthrn}_{1}(\alpha_{z})=0$, condition ${\mcal{R}_{1}^{1}(\alpha_{z})\geq \mcal{R}^{\{2,\stthrn\}}_{2}(\alpha_{z})}$ holds. Hence, after substituting $P_{1}^{\sttwon}(\alpha_{w})=P_{1}^{\stthrn}(\alpha_{w})=0$ in (\ref{eqn:S_fun}) and using the relation $\mcal{R}^{\{\stonen,\stnn\}}_{3}(\alpha_{w})=(1-\alpha_{w})\mcal{R}^{\{\stonen,\stnn\}}_{2}(\alpha_{w})$ from (\ref{eqn:proj}), we get:
\begin{IEEEeqnarray*}{rCl}
	\begin{array}{lll}
		\tau_{1}(\alpha_{w},1,0,0)\hspace{-0.2cm}&=&\hspace{-0.2cm} \mcal{R}_{1}^{1}(\alpha_{w})\\
		
		\tau_{2}(\alpha_{w},1,P^{\stonen}_{2},1)\hspace{-0.2cm}&=&\hspace{-0.2cm} \mcal{R}_{2}^{\{2,\stthrn\}}(\alpha_{w})+ P^{\stonen}_{2}(\alpha_{w}) \mcal{R}^{\{\stonen,\stnn\}}_{2}(\alpha_{w})\\
		
		\tau_{3}(\alpha_{w},0,P^{\stonen}_{2})\hspace{-0.2cm}&=&\hspace{-0.2cm}\mcal{R}_{3}^{\{3,\stonen,\sttwon,\stnn\}}(\alpha_{w})-(1-\alpha_{w})P^{\stonen}_{2}(\alpha_{w}) \mcal{R}^{\{\stonen,\stnn\}}_{2}(\alpha_{w}).
	\end{array}
\end{IEEEeqnarray*}
After balancing the buffer, we get the expressions for  $P_{2}^{\stonen}(\alpha_{w})$ and $\tau_{i}$ as listed in the Table. The listed expressions for 3b and 3c can be proved in a similar fashion.
\par\emph{\noindent {\bf\emph Remark 3:} It is clear from the Table-\ref{tab:usecasecointosstau} that in case 3b, the buffered throughput is more than $\mcal{R}_{1}^{1}(\alpha_{w})$ and $\mcal{R}_{2}^{2}(\alpha_{w})$ with direct throughput $\mcal{R}^{\{3,\stonen,\sttwon,\stnn\}}_{3}(\alpha_{z})$,  if $P_{1}^{\stthrn}(\alpha_{w})$ can balance out the buffer. Otherwise, depending on $P_{2}^{\stonen}(\alpha_{w})$ or $P_{1}^{\sttwon}(\alpha_{w})$ used, the buffered throughput is either $\mcal{R}_{1}^{1}(\alpha_{w})$ or $\mcal{R}_{2}^{2}(\alpha_{w})$, with direct throughput less than $\mcal{R}^{\{3,\stonen,\sttwon,\stnn\}}_{3}(\alpha_{z})$.}
\subsubsection*{Joint CCDF of Link SNRs in Scheme-1 and Scheme-2}

As defined in (\ref{eqn:jointProb}) and (\ref{eqn:link_rate_domainset})  the joint probability of the rate combination $R_{1}^{k_{1}},\,R_{2}^{k_{2}}$ and $R_{3}^{k_{3}}$ being the maximum feasible rates for linsk-1, 2 and 3, i.e. $P_{R_{1}^{k_{1}}R_{2}^{k_{2}}R_{3}^{k_{3}}}$ is required for carrying out the throughput analysis. This probability depends on the joint CCDF of link SNRs, i.e. $F_{\gamma_{1},\gamma_{2},\gamma_{3}}^{c}(y_{1},y_{2},y_{3})$. In this subsection, we evaluate the joint probability of rate-triplet $P_{R_{1}^{k_{1}}R_{2}^{k_{2}}R_{3}^{k_{3}}}$ for scheme-1 and scheme-2. We first state the necessary statistics required for the formulating the joint probability of both the schemes. 
\par\emph{\noindent {\bf\emph Lemma 5:} The expressions for CCDF and PDF of instantaeous SNR of link-2, i.e., $F_{\gamma_{2}}^{c}(y_{2})$, $f_{\gamma_{2}}(y_{2})$ together with the joint CCDF of instantaneous SNRs of link-1 and 3, i.e., $F_{\gamma_{1},\gamma_{3}}^{c}(y_{1},y_{3})$ are given by:
\begin{IEEEeqnarray}{rcl}\label{eqn:CCDFPDFexprn}
\begin{array}{lll}
	\hspace{0.4in}F_{\gamma_{2}}^{c}(y_{2}) \hspace{-0.1in}&=&\hspace{-0.1in} e^{-\frac{y_{2}}{\lambda_{2}}}\Big\{1-p_{2}+\frac{p_{2}}{1+\frac{y_{2}}{\mu_{2}}} \Big\},\label{eqn:cdf12}\\
	\hspace{0.4in}f_{\gamma_{2}}(y_{2}) \hspace{-0.1in}&=&\hspace{-0.1in}\frac{1}{\lambda_{2}} e^{-\frac{y_{2}}{\lambda_{2}}}\Big\{1-p_{2}+ \frac{p_{2}}{1+\frac{y_{2}}{\mu_{2}}}+ \frac{p_{2}\lambda_{2}/\mu_{2}}{\left(1+\frac{y_{2}}{\mu_{2}}\right)^{2}} \Big\},\label{eqn:pdf12}\\
	\hspace{-0.0in}F_{\gamma_{1},\gamma_{3}}^{c}(y_{1},y_{3}) \hspace{-0.1in}&=&\hspace{-0.1in} e^{-\left(\frac{y_{1}}{\lambda_{1}}+\frac{y_{3}}{\lambda_{3}}\right)}\Big\{1-p_{1}+ \frac{p_{1}}{1+\frac{y_{1}}{\mu_{1}}+\frac{y_{3}}{\mu_{3}}} \Big\}. \label{eqn:cdf11}
\end{array}
\end{IEEEeqnarray}
Proof:} Please see Appendix B.$\blacksquare$
\subsubsection*{Joint Probability of Rate-Triplet for Scheme-1 and Scheme-2}
\par We next formulate the joint probability of rate triplet for scheme-1 and scheme-2. As mentioned earlier, the elements of the index set $\{k_{1},k_{2},k_{3}\}$ can take any value independently in scheme-1 due to the mutual independence of indicator functions (\ref{eqn:ChanIndi_1_3}) and (\ref{eqn:ChanIndi_2_s1}), which leads to the cubic rate constellations.
\par\emph{\noindent {\bf\emph Lemma 6:} The joint probability of rate triplet  $R_{1}^{k_{1}}R_{2}^{k_{2}}R_{3}^{k_{3}}$ in scheme-1 is expressed as follows:
\begin{IEEEeqnarray}{rCl}\label{eqn:Pklm}
	\begin{array}{lll}
		\hspace{-.4cm}P_{R_{1}^{k_{1}}R_{2}^{k_{2}}R_{3}^{k_{3}}}=
		\hspace{.1cm}\sum\limits_{ j_{1}\in\{0,1\}}\sum\limits_{j_{2}\in\{0,1\}}\sum\limits_{ j_{3}\in\{0,1\}}\hspace{0.3cm}(-1)^{j_{1}+j_{2}+j_{3}}F_{\gamma_{1},\gamma_{2},\gamma_{3}}^{c}(\gamma_{1}^{j_{1}+k_{1}},\gamma_{2}^{j_{2}+k_{2}},\gamma_{3}^{j_{3}+k_{3}}),\vspace{-0.0cm}\\
	\end{array}
\end{IEEEeqnarray}
where $F_{\gamma_{1},\gamma_{2},\gamma_{3}}^{c}(y_{1},y_{2},y_{3})$ is the joint CCDF of instantaneous SNRs of link 1, 2 and 3, which is given by (\ref{eqn:CCDFScheme1}) for scheme-1.\\
Proof:} Using  (\ref{eqn:ChanIndi_1_3}), (\ref{eqn:ChanIndi_2_s1}) and (\ref{eqn:jointProb}), we write the joint probability in terms of instantaneous SNR as:
\begin{IEEEeqnarray}{rCl}\label{eqn:jointProb2}
	\hspace{-.5cm}P_{R_{1}^{k_{1}}R_{2}^{k_{2}}R_{3}^{k_{3}}}&=&\Pr\{\gamma_{1}^{k_{1}}\leq\gamma_{1}<\gamma_{1}^{k_{1}+1},\gamma_{2}^{k_{2}}\leq\gamma_{2}<\gamma_{2}^{k_{2}+1},\gamma_{3}^{k_{3}}\leq\gamma_{3}<\gamma_{3}^{k_{3}+1}\}.
\end{IEEEeqnarray}
Now, after expanding the above equation using $\Pr\{\gamma_{i}^{k_{i}}\leq\gamma_{i}<\gamma_{i}^{k_{i}+1}\}=F_{\gamma_{i}}^{c}(\gamma_{i}^{k_{i}})-F_{\gamma_{i}}^{c}(\gamma_{i}^{1+k_{i}}),$ we get (\ref{eqn:Pklm}). Furthermore, it is clear from the Fig. \ref{fig:sysmod1} and equation (\ref{eqn:InsSNR}), that $g_{1}=g_{3}$ is common to both link-1 and link-3, which makes SNRs $\gamma_{1}$ and $\gamma_{3}$ dependent. Hence, after substituting the expressions of $F_{\gamma_{1},\gamma_{3}}^{c}(y_{1},y_{3})$ and $F_{\gamma_{2}}^{c}(y_{2})$ from (\ref{eqn:CCDFPDFexprn}) in $F_{\gamma_{1},\gamma_{2},\gamma_{3}}^{c}(y_{1},y_{2},y_{3})=F_{\gamma_{1},\gamma_{3}}^{c}(y_{1},y_{3})F_{\gamma_{2}}^{c}(y_{2})$, we get (\ref{eqn:CCDFScheme1}).$\blacksquare$
\begin{figure*}[!t]
\hrulefill
\begin{IEEEeqnarray}{rCl}\label{eqn:CCDFScheme1}
	\hspace{-0.5cm}F_{\gamma_{1},\gamma_{2},\gamma_{3}}^{c}(y_{1},y_{2},y_{3}) \hspace{-0.1cm}&=&\hspace{-0.1cm} e^{-\left(\frac{y_{1}}{\lambda_{1}}+\frac{y_{2}}{\lambda_{2}}+\frac{y_{3}}{\lambda_{3}}\right)}\hspace{-0.1cm}\left[1-p_{1}+\frac{p_{1}} {1+\frac{y_{1}}{\mu_{1}}+\frac{y_{3}}{\mu_{3}}}\right]\hspace{-0.2cm}\left[1-p_{2}+\frac{p_{2}}{1+\frac{y_{2}}{\mu_{2}}} \right]\hspace{-0.15cm};\text{In scheme-1.}
\end{IEEEeqnarray}
\hrulefill
\hrulefill
\end{figure*}
\par Substituting (\ref{eqn:CCDFScheme1}) in (\ref{eqn:Pklm}), we get the closed form expression of joint probability in scheme-1. For scheme-2, the  elements of a index-set $\{k_{1},k_{2},k_{3}\}$ for which $k_{2}<k_{3}$, are restricted  (the probability of occurrence of such events is zero) due to dependence of $I_{u_{2}}^{k_{2}}$ on $I_{u_{3}}^{k_{3}}$ given by (\ref{eqn:ChanIndi_1_3}) and (\ref{eqn:ChanIndi_2_s2}), , which leads to the prism rate constellations. 
 The probability of occurrence of index $k_{2}$ for $k_{2}=k_{3}$ increases due to the enhancement of the probability activation/partition region for $\gamma_{2}$ from $\gamma_{2}^{k_{2}}\leq \gamma_{2}\leq\gamma_{2}^{k_{2}+1}$ to $0\leq \gamma_{2}\leq\gamma_{2}^{k_{2}+1}$. Hence, it is clear that $j_{2}$ of $\gamma_{2}^{j_{2}+k_{2}}$ in (\ref{eqn:Pklm}) is not $j_{2}=0$ but $j_{2}=-k_{2}$ when $k_{2}=k_{3}$, and in sign-flip argument, $j_{2}$ is replaced by $\max(j_{2},0)$ to maintain consistency.
\par\emph{\noindent {\bf\emph Lemma 7:}  The joint probability of rate index $R_{1}^{k_{1}}R_{2}^{k_{2}}R_{3}^{k_{3}}$ in scheme-2 is expressed as follows:
\begin{IEEEeqnarray}{rCl}\label{eqn:Pklm2}
	\begin{array}{lll}
		\hspace{-.4cm}P_{R_{1}^{k_{1}}R_{2}^{k_{2}}R_{3}^{k_{3}}}=
		\hspace{.1cm}\sum\limits_{ j_{1}\in\{0,1\}}\sum\limits_{j_{2}\in\mcal{I}(k_{2},k_{3})}\sum\limits_{ j_{3}\in\{0,1\}}\hspace{0.3cm}(-1)^{j_{1}+\max(j_{2},0)+j_{3}}F_{\gamma_{1},\gamma_{2}+\gamma_{3},\gamma_{3}}^{c}(\gamma_{1}^{j_{1}+k_{1}},\gamma_{2}^{j_{2}+k_{2}},\gamma_{3}^{j_{3}+k_{3}}),\vspace{-0.0cm}\\
	\end{array}
\end{IEEEeqnarray}
where $j_{2}$ takes value over the integer set $\mcal{I}(k_{2},k_{3})$, which depends on $k_{2}$ and $k_{3}$, and is given as:
\begin{IEEEeqnarray}{rCl}\label{eqn:I_k}
	\mcal{I}(k_{2},k_{3})=\left\{
\begin{array}{lll}
	\{\phi\}\quad  \text{if } k_{2}<k_{3}\\
	\{-k_{2},1\}\quad \text{if } k_{2}=k_{3}\\
	\{0,1\}\quad \text{if } k_{2}>k_{3},
\end{array}\right.
\end{IEEEeqnarray}
where $F_{\gamma_{1},\gamma_{2},\gamma_{3}}^{c}(y_{1},y_{2},y_{3})$ is the joint CCDF of instantaneous SNRs of links 1, 2 and 3, which is given by (\ref{eqn:CCDFScheme2}) for scheme-2.\\
Proof:} The joint CCDF for scheme-2, i.e. $F_{\gamma_{1},\gamma_{2}+\gamma_{3},\gamma_{3}}^{c}(y_{1},y_{2},y_{3})$ is expressed as follows:
\begin{figure*}[!t]
	\hrulefill
	\begin{IEEEeqnarray}{rCl}\label{eqn:CCDFScheme2}
	\hspace{-0.2cm}F_{\gamma_{1},\gamma_{2},\gamma_{3}}^{c}(y_{1},y_{2},y_{3})\hspace{-0.1cm}&=&\hspace{-0.1cm} \underbrace{F_{\gamma_{1},\gamma_{3}}^{c}(y_{1},y_{3})F_{\gamma_{2}}^{c}(y_{4})}_\text{I}-\underbrace{F_{\gamma_{2}}(y_{2})F_{\gamma_{1},\gamma_{3}}^{c}(y_{1},y_{2})}_\text{II}+\underbrace{\int\limits_{0}^{y_{4}}F_{\gamma_{1},\gamma_{3}}^{c}(y_{1},y_{2}-x)f_{\gamma_{2}}(x)dx}_\text{III};\nonumber\\
		&&\hspace{4.5cm}\text{where }y_{4}=\max((y_{2}-y_{3}), 0) \text{ In scheme-2}, 
	\end{IEEEeqnarray}
	\hrulefill
	\begin{IEEEeqnarray}{rCl}\label{eqn:CCDFScheme3}
		\begin{array}{lll}
			\hspace{-0.25cm} F_{\gamma_{1},\gamma_{2},\gamma_{3}}^{c}(y_{1},y_{2},y_{3}) &\overset{\text{PIP}}{\equiv}& \left(1+\frac{y_{1}}{\mu_{1}}+\frac{y_{3}}{\mu_{3}}\right)^{-1}\left(1+\frac{y_{4}}{\mu_{2}}\right)^{-1}- \left(1+\frac{y_{1}}{\mu_{1}}+\frac{\max(y_{2},y_{3})}{\mu_{3}}\right)^{-1}\\
			\hspace{-0.25cm}&&\hspace{-4.3cm}\times\left[1-\left(1+\frac{y_{2}}{\mu_{2}}\right)^{-1} \right]\frac{\frac{y_{4}}{\mu_{2}}}{(1+\frac{y_{4}}{\mu_{2}})(1+\frac{y_{1}}{\mu_{1}}+\frac{y_{2}+\mu_{2}}{\mu_{3}})}+\frac{\frac{\mu_{2}}{\mu_{3}}}{(1+\frac{y_{1}}{\mu_{1}}+\frac{y_{2}+\mu_{2}}{\mu_{3}})^2}\log\left[(1+\frac{\frac{y_{4}}{\mu_{2}}}{\frac{\mu_{3}}{\mu_{2}}(1+\frac{y_{1}}{\mu_{1}}+\frac{y_{3}}{\mu_{3}})})(1+\frac{y_{4}}{\mu_{2}})\right]\hspace{-0.1cm};\\
			&&\hspace{1.1cm}\text{where }y_{4}=\max((y_{2}-y_{3}), 0) \text{ for PIP case, in scheme-2}.
		\end{array}
	\end{IEEEeqnarray}
	\hrulefill
\end{figure*} 
\begin{figure}[t] 
	\begin{center}
		\includegraphics[scale=0.6]{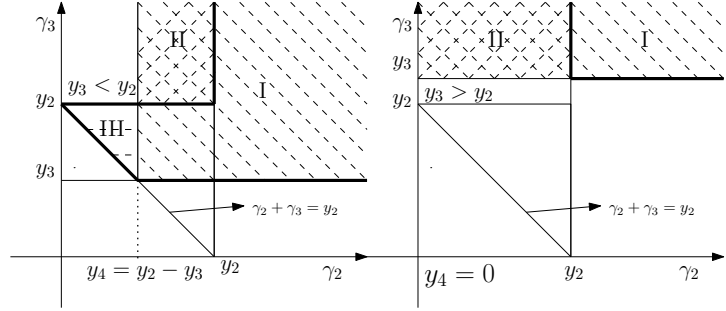}
	\end{center}
	\vspace{-0.2cm}
	\caption{SNR region of $\gamma_{2}$ and $\gamma_{3}$ given $\gamma_{1}\geq y_{1}$}
	\label{fig:SNR_REGION}
\end{figure}
\begin{IEEEeqnarray*}{rCl}
	\begin{array}{lll}
		F_{\gamma_{1},\gamma_{2}+\gamma_{3},\gamma_{3}}^{c}(y_{1},y_{2},y_{3})&=& 
		\Pr\{\gamma_{1}\geq y_{1},\,\gamma_{2}+\gamma_{3}\geq y_{2},\,\gamma_{3}\geq y_{3}\}\\
		&=&\Pr\{\gamma_{1}\geq y_{1},\,\gamma_{2}\geq y_{2}-\gamma_{3},\,\gamma_{3}\geq y_{3}\},
	\end{array}
\end{IEEEeqnarray*}
which can be broken in two parts, i.e. $y_{2}>y_{3}$ and $y_{2}\leq y_{3}$, as follows:
\begin{IEEEeqnarray}{rCl}
\begin{array}{lll}
F_{\gamma_{1},\gamma_{2}+\gamma_{3},\gamma_{3}}^{c}(y_{1},y_{2},y_{3})=\left\{
\begin{array}{ll}
\Pr\{\gamma_{1}\geq y_{1},\,\gamma_{2}\geq y_{2}-\gamma_{3},\,\gamma_{3}\geq y_{3}\}\ \   \text{if } y_{2}>y_{3}\\
\Pr\{\gamma_{1}\geq y_{1},\,\gamma_{2}\geq 0,\,\gamma_{3}\geq y_{3}\}\ \qquad\quad  \text{if } y_{2}=y_{3},\\
0\ \hspace{6.2cm}  \text{if } y_{2}<y_{3}.
\end{array}\right.
\end{array}
\end{IEEEeqnarray}
When complete adaptive (continuous) rate transmission is used, combined $\gamma_{2}+\gamma_{3}$ is always superior to $\gamma_{3}$. But, with discrete rate transmission considered here, $\gamma_{2}+\gamma_{3}$ might not result in higher rate than $\gamma_{3}$. Also, much of the advantage of direct path is captured by rate and link selection, and as will be shown in the next section, combining offers very little additional throughput. After defining $y_{4}=\max(y_{2}-y_{3},0)$, the resultant CCDF can be broken in two parts. Fig. \ref{fig:SNR_REGION} indicates the  SNR regions of $\gamma_{2}$ and $\gamma_{3}$ given  $\gamma_{1}\geq y_{1}$ for the two regions i.e. $y_{4}>0$ and $y_{4}=0$. It is clear from this figure that when $y_{3}>y_{2}$,  CCDF of both the schemes are the same. Hence using Fig. \ref{fig:SNR_REGION}, the CCDF of link SNRs  with scheme-2 is given by (\ref{eqn:CCDFScheme2}).$\blacksquare$
\par After substituting the expressions for $F_{\gamma_{1},\gamma_{3}}^{c}(y_{1},y_{3})$ and $F_{\gamma_{2}}^{c}(y_{2})$ from (\ref{eqn:CCDFPDFexprn}) in (\ref{eqn:CCDFScheme2}), we expand the expression. The integral  in (\ref{eqn:CCDFScheme2}) can be expressed in closed form, but  is omitted due to paucity of space. We however present the closed-form expression for PIP case in (\ref{eqn:CCDFScheme3}) (when $\lambda_{1},\lambda_{2},\lambda_{3}\rightarrow\infty$ and $p_{1}=p_{2}=1$).

\section{Numerical Example and Simulations}
In this section, we evaluate the throughput performance by simulation, and compare the same with the derived analytical expressions.

\par We first show using a numerical example that scheme-2 is beneficial only when link-2 is weak. Table-\ref{tab:ProbDomainsetsinglerate} lists the probabilities of selection of various modes for single equal rate at the source and relay. The rates are set as $R_{1}^{1}=R_{2}^{1}=2$ in the PIP regime where $\gamma_{p}=-5$ dB. The normalized distances used are  $d_{1}=d_{2}=1$, $d_{3}=2$ and $d_{1p}=3$. Now for comparison purpose, we assume the primary is relatively close to relay i.e. $d_{2p}=1.5$ in the first scenario than in the second scenario, in which $d_{2p}=3.0$.
{\normalsize
	\begin{table}[h]
		\caption{Joint Probabilities associated with domain sets for equal single rate}
		\footnotesize
		\renewcommand{\arraystretch}{1.2}
		\label{tab:ProbDomainsetsinglerate} 
		\begin{minipage}{\textwidth}
					\centering
			\begin{tabular}{|p{.5cm}|p{1.0cm}|p{.85cm}|p{.85cm}|p{.85cm}|p{.85cm}|p{.85cm}|p{.85cm}|p{.5cm}|p{.85cm}||p{1.2cm}| }
				\hline
				&\multicolumn{9}{c}{ $R_{1}^{1}=R_{2}^{2}=2,\,\gamma_{p}=-5$ dB, $d_{1}=d_{2}=1$, $d_{3}=2$,   $d_{1p}=3$, and $d_{2p}=1.5$.}\\
				\hline
				$\alpha_{w}$& Scheme &  $\mcal{U}^{1}$ & $\mcal{U}^{2}$ & $\mcal{U}^{3}$ &  $\mcal{U}^{\stonen}$ & $\mcal{U}^{\sttwon}$ & $\mcal{U}^{\stthrn}$  & $\mcal{U}^{\stnn}$ & $\mcal{U}^{\N}$ & $\tau_{t}(\alpha_{w})/2$ \\\hline
				
				$\alpha_{0}$ & 1 & \hspace{0.05cm}$0$ & $\mbf{0.1935}$ & $0.1935$  & $0.0689$ & \hspace{0.05cm}$0$ & \hspace{0.05cm}$0$  &  \hspace{0.05cm}$0$ &$0.5440$ & \hspace{0.05cm}$0.4559$\\\hline
				
				$\alpha_{0}$ & 2 & \hspace{0.05cm}$0$ & $\mbf{0.2689}$ & \hspace{0.05cm}$0$  & $0.2624$ & \hspace{0.05cm}$0$ & \hspace{0.05cm}$0$  &  \hspace{0.05cm}$0$ &$0.4687$ & \hspace{0.05cm}$0.5313$\\\hline\hline
				
				$\alpha_{1}$ & 1 & $\mbf{0.3686}$ & $\mbf{0.0624}$ & $\mbf{0.2624}$   & \hspace{0.05cm}$0$ & \hspace{0.05cm}$0$ & $0.1311$   &  \hspace{0.05cm}$0$ &$0.1754$ & \hspace{0.05cm}$0.5435$ \\\hline
				
				$\alpha_{1}$ & 2 & $\mbf{0.3105}$ & $\mbf{0.0797}$ & $\mbf{0.2624}$   & \hspace{0.05cm}$0$ & \hspace{0.05cm}$0$ & $0.1892$   &  \hspace{0.05cm}$0$ &$0.1582$ & \hspace{0.05cm}$0.5521$ \\\hline\hline
				
				$\alpha_{2}$ & 1 & $\mbf{0.4997}$ & \hspace{0.11cm}$0$ & $0.0221$ & \hspace{0.11cm}$0$  & $0.2403$ & \hspace{0.11cm}$0$ &  \hspace{0.05cm}$0$ &$0.2379$   & \hspace{0.05cm}$0.7621$  \\\hline
				
				$\alpha_{2}$ & 2 & $\mbf{0.4997}$ & \hspace{0.11cm}$0$ & $0.0221$ & \hspace{0.11cm}$0$  & $0.2403$ & \hspace{0.11cm}$0$ &  \hspace{0.05cm}$0$ &$0.2379$   & \hspace{0.05cm}$0.7621$  \\\hline\hline

				&\multicolumn{9}{c}{ $R_{1}^{1}=R_{2}^{2}=2,\,\gamma_{p}=-5$ dB, $d_{1}=d_{2}=1$, $d_{3}=2$,   $d_{1p}=3$, and $d_{2p}=3$.}\\
				\hline
				$\alpha_{w}$& Scheme &  $\mcal{U}^{1}$ & $\mcal{U}^{2}$ & $\mcal{U}^{3}$ &  $\mcal{U}^{\stonen}$ & $\mcal{U}^{\sttwon}$ & $\mcal{U}^{\stthrn}$  & $\mcal{U}^{\stnn}$ & $\mcal{U}^{\N}$ & $\tau_{t}(\alpha_{w})/2$ \\\hline
				
				$\alpha_{0}$ & 1 & \hspace{0.05cm}$0$ & $\mbf{0.5458}$ & $0.0682$  & $0.1942$ & \hspace{0.05cm}$0$ & \hspace{0.05cm}$0$  &  \hspace{0.05cm}$0$ &$0.1918$ & \hspace{0.05cm}$0.8082$\\\hline
				
				$\alpha_{0}$ & 2 & \hspace{0.05cm}$0$ & $\mbf{0.5936}$ & \hspace{0.05cm}$0$  & $0.2624$ & \hspace{0.05cm}$0$ & \hspace{0.05cm}$0$  &  \hspace{0.05cm}$0$ &$0.1440$ & \hspace{0.05cm}$0.8560$\\\hline\hline
				
				$\alpha_{1}$ & 1 & $\mbf{0.1299}$ & $\mbf{0.1760}$ & $\mbf{0.2624}$   & \hspace{0.05cm}$0$ & \hspace{0.05cm}$0$ & $0.3698$   &  \hspace{0.05cm}$0$ &$0.0618$ & \hspace{0.05cm}$0.6003$ \\\hline
				
				$\alpha_{1}$ & 2 & $\mbf{0.0939}$ & $\mbf{0.1878}$ & $\mbf{0.2624}$   & \hspace{0.05cm}$0$ & \hspace{0.05cm}$0$ & $0.4058$   &  \hspace{0.05cm}$0$ &$0.0501$ & \hspace{0.05cm}$0.6062$ \\\hline\hline
				
				$\alpha_{2}$ & 1 & $\mbf{0.4997}$ & \hspace{0.11cm}$0$ & $0.0221$ & \hspace{0.11cm}$0$  & $0.2403$ & \hspace{0.11cm}$0$ &  \hspace{0.05cm}$0$ &$0.2379$   & \hspace{0.05cm}$0.7621$  \\\hline
				
				$\alpha_{2}$ & 2 & $\mbf{0.4997}$ & \hspace{0.11cm}$0$ & $0.0221$ & \hspace{0.11cm}$0$  & $0.2403$ & \hspace{0.11cm}$0$ &  \hspace{0.05cm}$0$ &$0.2379$   & \hspace{0.05cm}$0.7621$  \\\hline\hline

			\end{tabular}
			\vspace{-0.175cm}
			\label{tab:CCDF_LSP} 
		\end{minipage}
	\end{table}
} 
It is clear from  Table-\ref{tab:ProbDomainsetsinglerate} that when scheme-2 is used, the probability of silence intervals (mode $\mcal{U}^{\N}$) decreases for $\alpha_{0}$ and $\alpha_{1}$. Also, the direct path is not affected by scheme-2. Since according to stability condition $\mcal{R}_{2}^{2}(\alpha_{0})<\mcal{R}_{1}^{1}(\alpha_{1})$ for both the schemes,  the buffer is stable for $\alpha_{0}$, which is also clear by looking at the minimum of $\tau_{t}(\alpha_{w})$. It is clear from the table that as $\alpha_{w}$ decreases, the system has fewer silent intervals with scheme-2 as compared to scheme-1. Hence, the advantage of scheme-2 will be more pronounced when link-2 is heavily attenuated. The additional advantage due to combining is minimal for larger value of $\alpha_{w}$ in CRN, and most throughput gains are attained due to link and rate selection itself. When link-2 is not heavily attenuated, there is very little gain in throughput with use of scheme-2 as the buffer already underflows for lower $\alpha_{w}$.
\subsubsection*{Simulation}
\par
\begin{figure*}[t]
	\vspace*{-0.0cm}
	\begin{multicols}{3}		
		\includegraphics[height=5.cm,  width=0.33\textwidth]{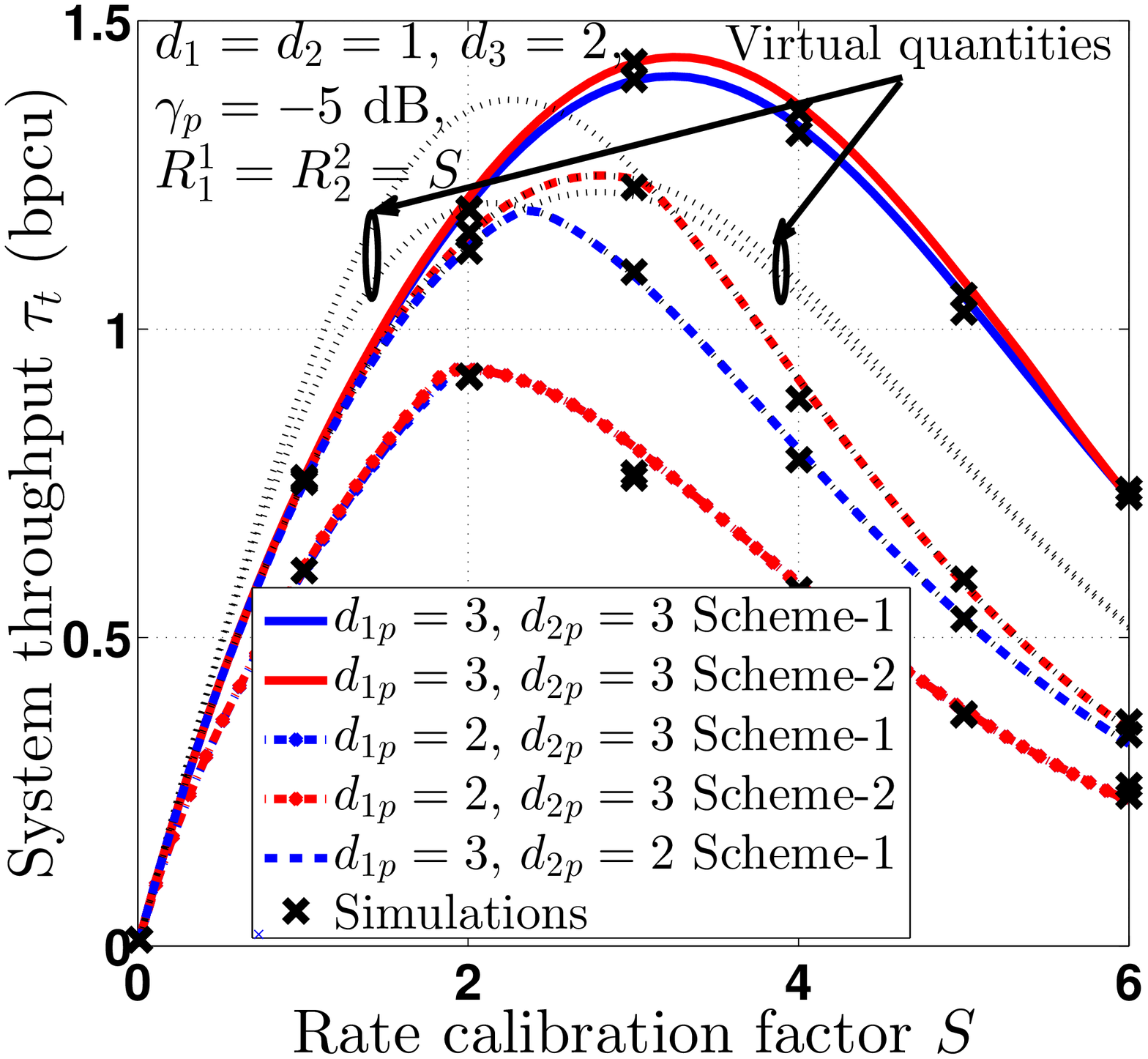}
		\par\caption{System throughput vs. $S$ of Scheme-1 and 2 with fixed rate  in PIP regime for different $d_{1p}$, $d_{2p}$}\label{fig:work3_fig2}

		\includegraphics[height=5.cm,  width=0.33\textwidth]{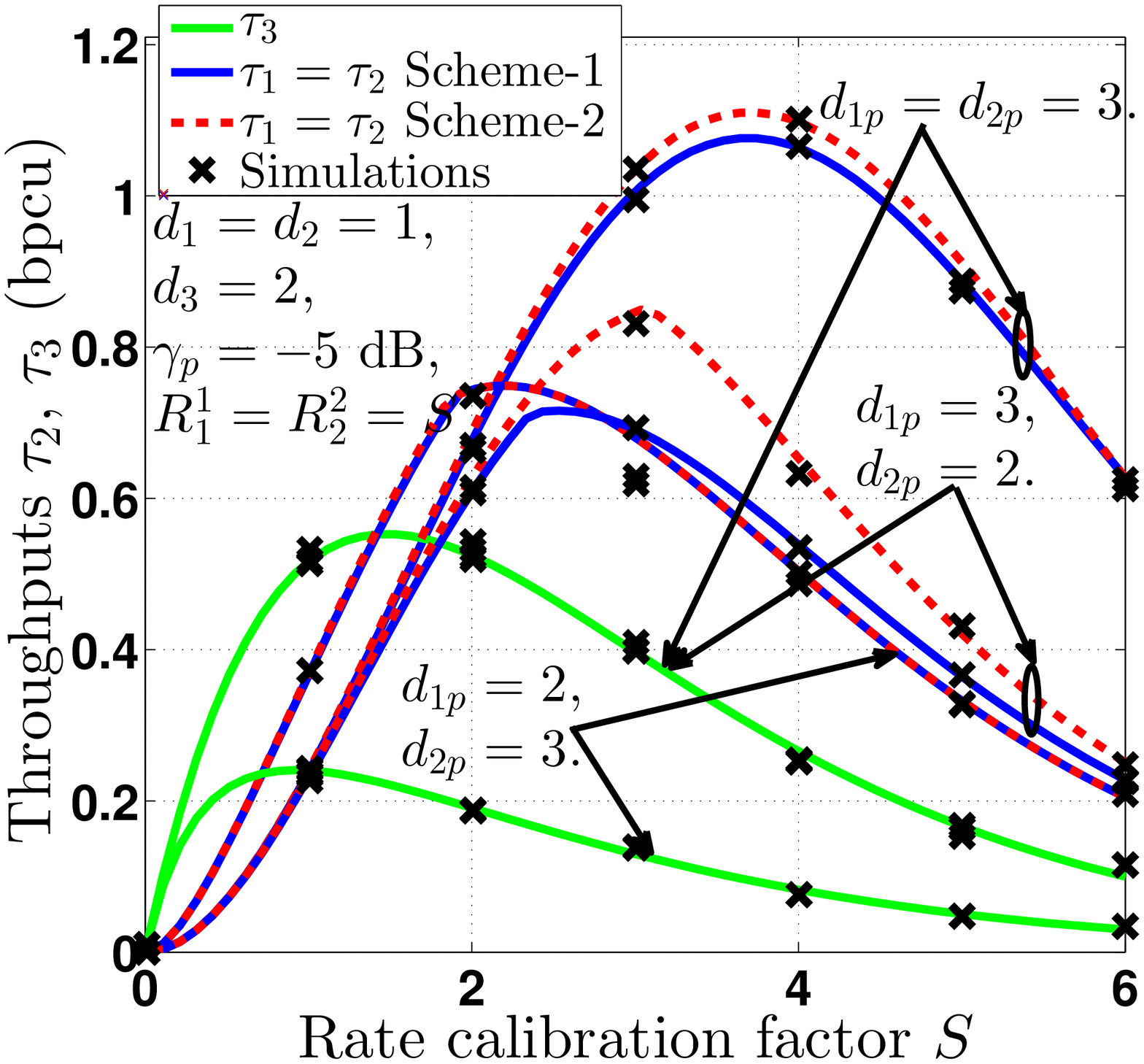}
		\par\caption{Buffered/Direct throughput vs. $S$  of Scheme-1 and 2 with fixed rate and different schemes in PIP regime for different $d_{1p}$, $d_{2p}$}\label{fig:work3_fig3}

		\includegraphics[height=5.cm,  width=0.33\textwidth]{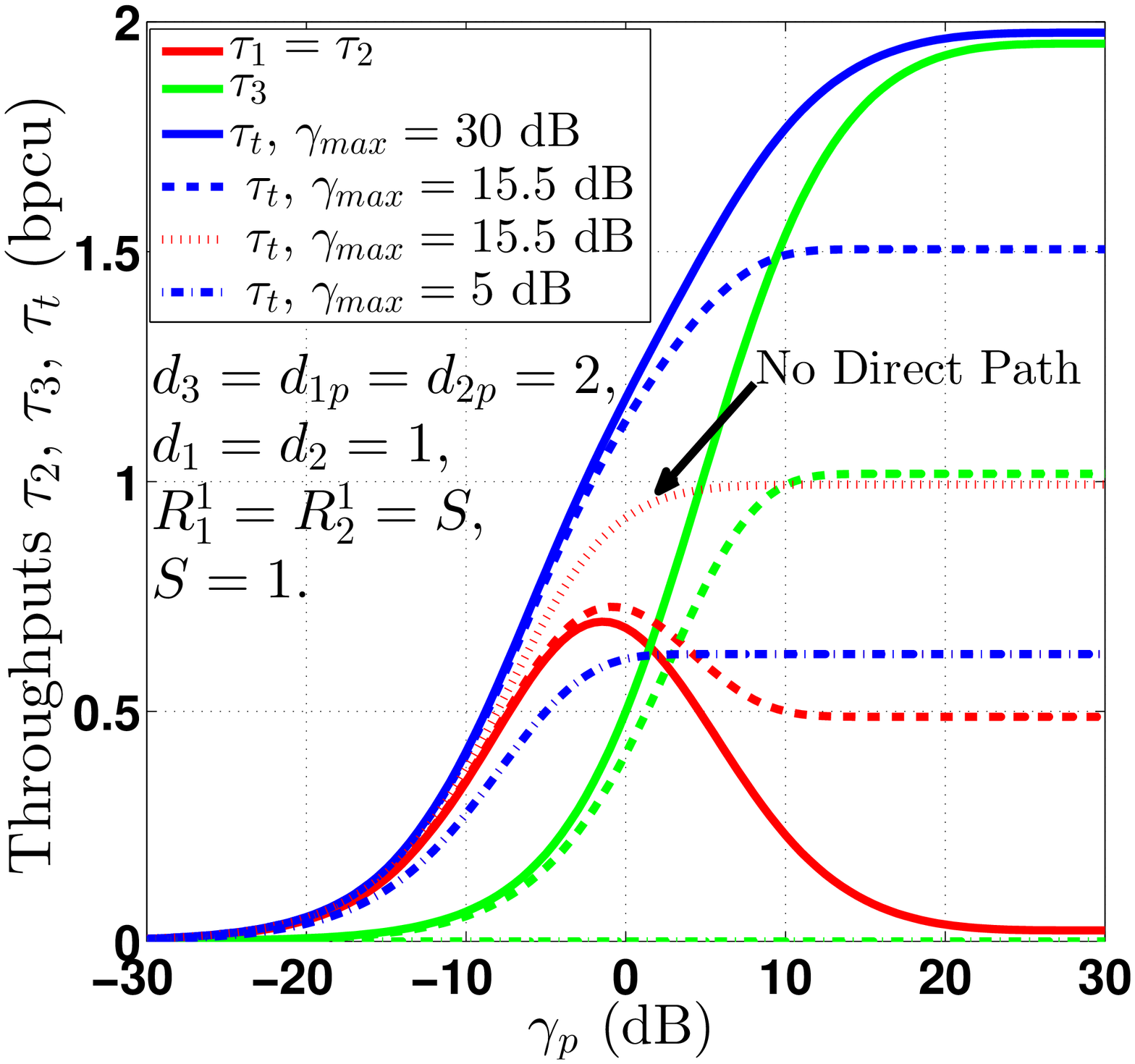}
		\par\caption{Throughputs vs. $\gamma_{p}$ of Scheme-1 for fixed rate and different value of $\gamma_{max}$} \label{fig:work3_fig1}
	\end{multicols}
	\begin{multicols}{3}
		\includegraphics[height=5.cm,  width=0.33\textwidth]{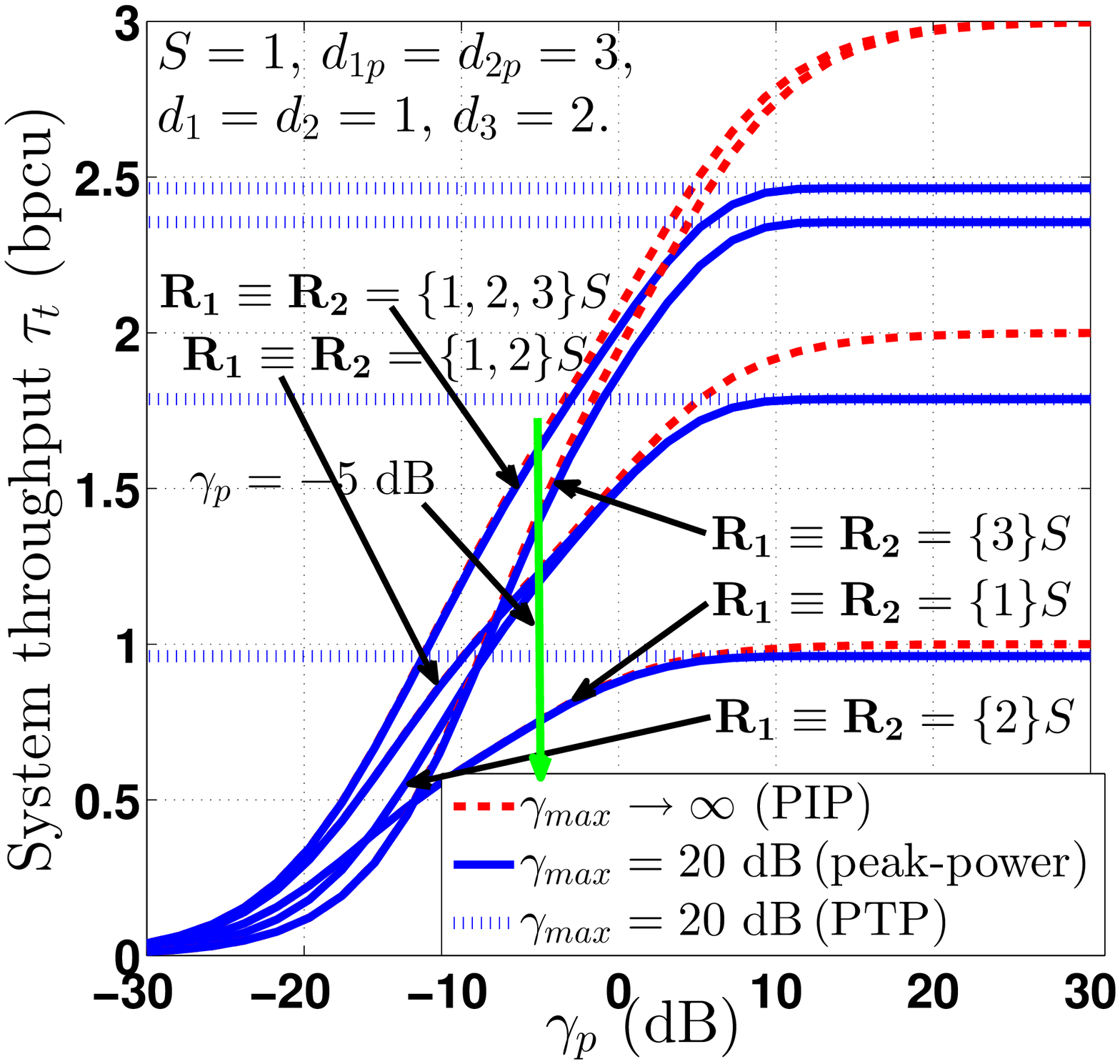}
		\caption{System throughput vs. $\gamma_{p}$  of scheme-1 with $S=1$ for different discrete rates and $\gamma_{max}$.}\label{fig:work3_fig4}

		\includegraphics[height=5.cm,  width=0.33\textwidth]{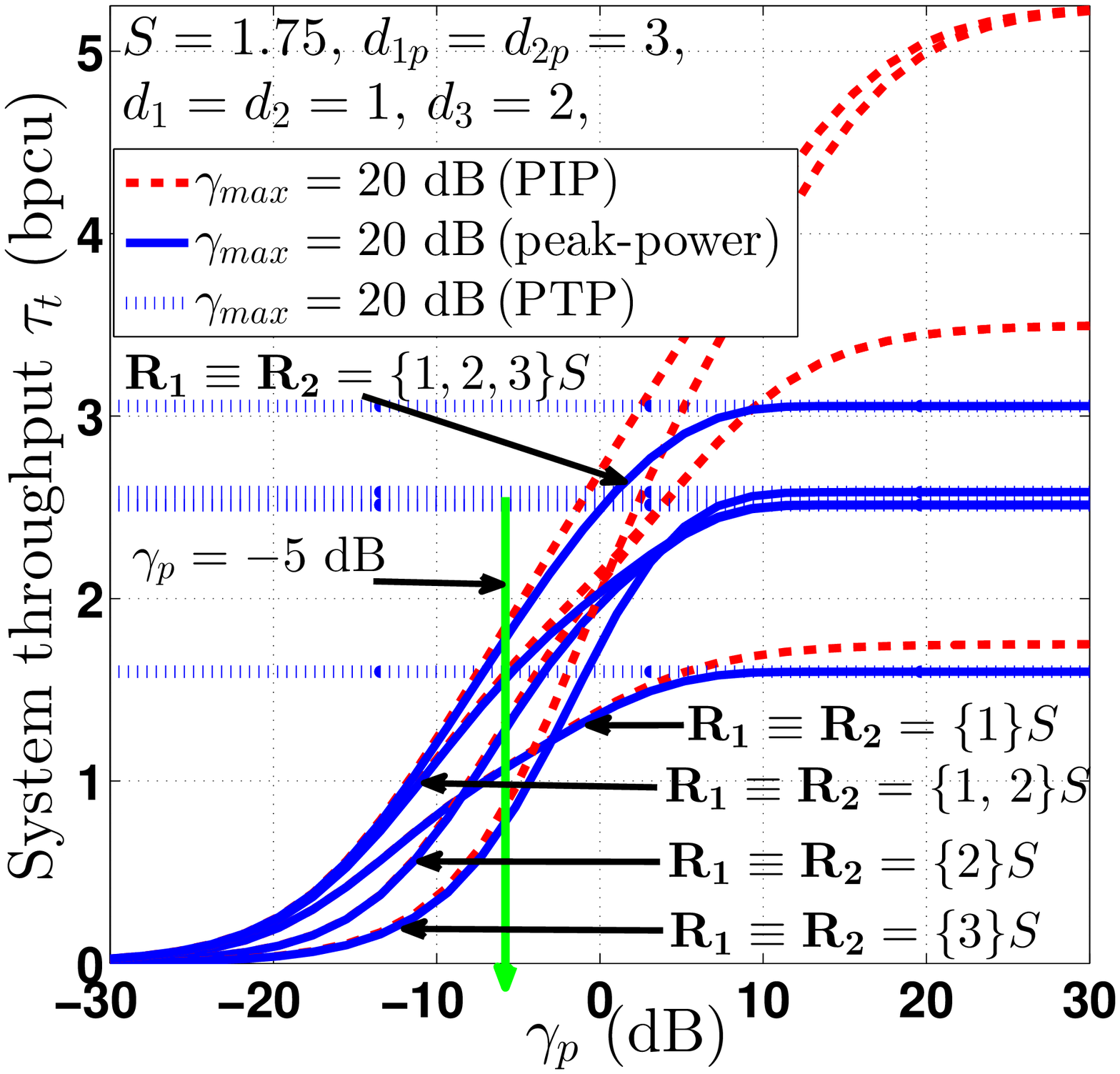}
		\caption{System throughput vs. $\gamma_{p}$  of scheme-1 with $S=1.75$ for different discrete rates and $\gamma_{max}$.}\label{fig:work3_fig5}

		\includegraphics[height=5.cm,  width=0.33\textwidth]{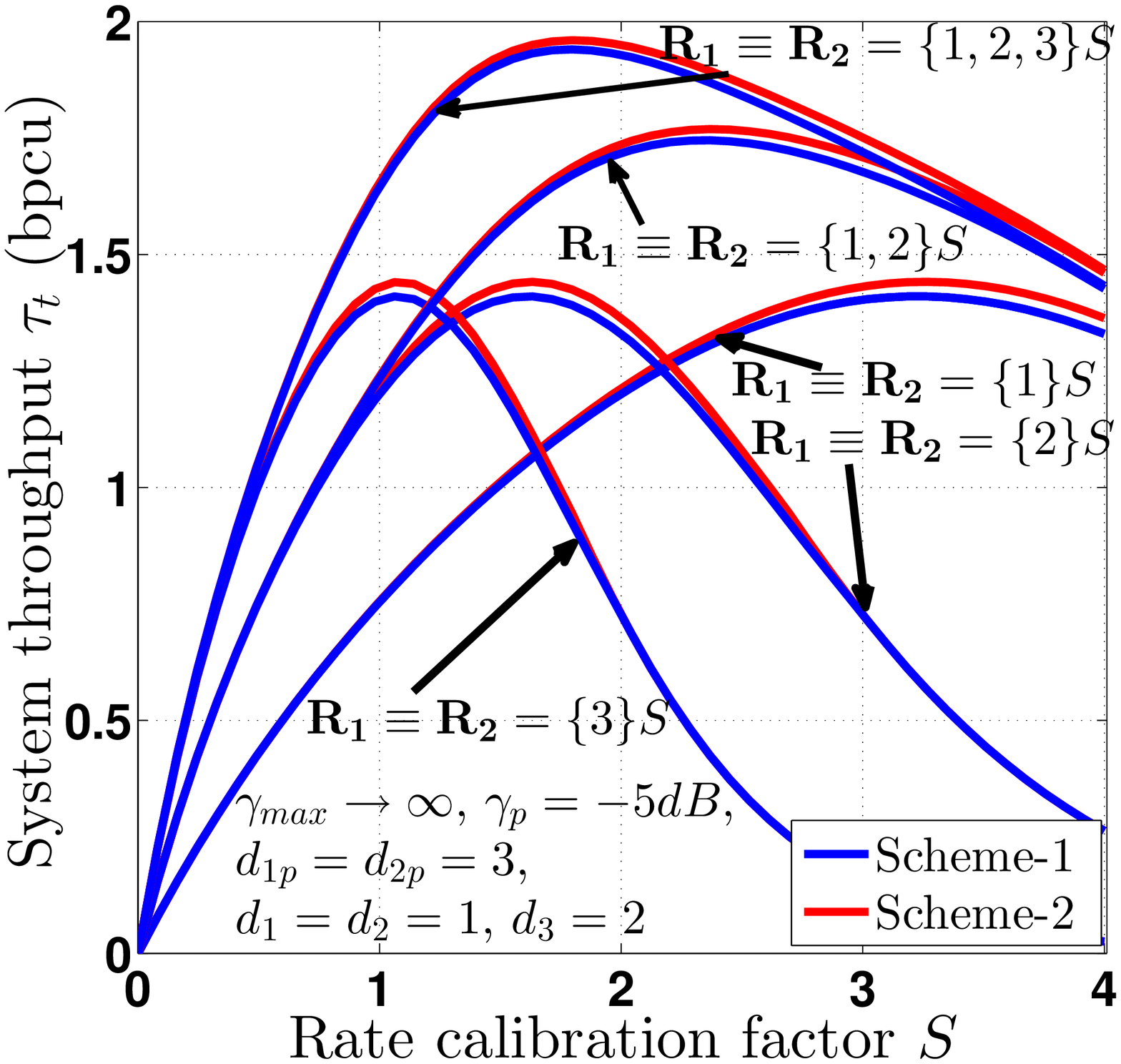}
		\caption{System throughput vs. $\gamma_{p}$  of schemes-1 and 2  in PIP regime with $\gamma_{p}=-5$ dB for different discrete rates.}\label{fig:work3_fig6}
		\vspace{-0.25 in}
	\end{multicols}
	\vspace*{-0.85cm}
\end{figure*}

 For Fig. \ref{fig:work3_fig2}, \ref{fig:work3_fig3} and \ref{fig:work3_fig1},  $R_{1}^{1}=R_{1}^{2}=S=1$ (integer $S$),  we set distances $d_{1}=d_{2}=1,\,d_{3}=2$.
\par Fig. \ref{fig:work3_fig2} and Fig. \ref{fig:work3_fig3} compare the performance of scheme-1 and 2 in PIP case when $\gamma_{p}=-5$ dB. Fig. \ref{fig:work3_fig2} depicts the system throughput for different combination of $d_{1p}$ and $d_{2p}$, whereas Fig. \ref{fig:work3_fig3} depicts the buffered and direct throughput related to \ref{fig:work3_fig2}. It is observed in Fig. \ref{fig:work3_fig2} that system throughput is not always a differentiable function of the rate calibration factor $S$. This is because of the switch between different regions that arises due to discrete rates. It is clear from the Fig. \ref{fig:work3_fig2} that scheme-2 does not always result in substantially larger throughput than scheme-1. More insight is obtained from the Fig.~\ref{fig:work3_fig3}, where it is observed that when source is close to primary, direct as well as buffered throughputs are small, and hence there is little difference between performance of scheme-1 and scheme-2. On the other hand, when link-2 is weak, scheme-2 results in much better performance than scheme-1. The difference between the two schemes is higher for larger $S$, which is evident from Fig. \ref{fig:work3_fig2}. It was shown in Fig. \ref{fig:work3_fig2} and \ref{fig:work3_fig3} that the derived expressions are accurate, and perfectly match with the simulation results. Extensive computer simulations have shown that the derived expressions are accurate for all system parameters. To ensure clarity,  we omit the simulation plots in subsequent figures.
\par Fig.~\ref{fig:work3_fig1} shows the throughput vs $\gamma_{p}$ in both PTP and PIP regimes with $d_{2p}=d_{rp}=2$ for various $\gamma_{max}$ when  $R_{1}^{1}=R_{2}^{1}=1$. The throughput is plotted for peak SNR $\gamma_{max}$ of  $30, 10.6$ and $0$ dB. The throughput for the case when the direct path is shadowed is also plotted for  $\gamma_{max}=10.6$ dB. It is clear from these plots that the direct path is almost always picked in high-SNR scenarios. In other scenarios, the role of relay and its buffer becomes apparent. In other words, under fixed statistics and SNR, the direct path is picked for lower rates, which minimises the usage of buffer, whereas the relayed path is used more often at  higher rates.
\par For Fig. \ref{fig:work3_fig4}, \ref{fig:work3_fig5} and \ref{fig:work3_fig6},  we set symmetric distances $d_{1}=d_{2}=1,\,d_{3}=2,\,d_{1p}=d_{2p}=3$. Fig. \ref{fig:work3_fig4} and Fig. \ref{fig:work3_fig5} depict the throughput performace versus $\gamma_{p}$ in scheme-1 assuming discrete rates with $S=1$ and $S=1.75$. It is apparent from these figures that in high-SNR (low-SNR) scenario, the  throughput is mainly due to selection of the largest (smallest) rate. In the medium-SNR regime, the contribution of all rates is evident. Adding more discrete rates is not going to increase the throughput at low and medium SNRs. 
\par Fig. \ref{fig:work3_fig6} depicts the throughput performace versus $S$ of scheme-1 and 2 assuming discrete rates in the PIP regime, when $\gamma_{p}=-5$ dB. The importance of using discrete rates is apparent at medium and high SNRs. The contribution of scheme-2 is minimal as link-2 is not weak.

\section*{Conclusion}
In this paper, we presented analysis of performance of a buffered DF relay based three-node underlay cooperative cognitive relay network with a direct path. We assumed use of multiple rates at the source and the relay. We performed joint link and rate selection. It was shown that combining the signal from the source and the relay does not improve performance except when the second hop is weak. Comprehensive analysis was presented that brought insights on buffer stability and throughput.
\section*{Appendix A}\vspace{-0.15cm}
\section*{Proof of Lemma-3}\vspace{-0.15cm}
As already discussed, we consider  three primary cases: case-1 when link-2 is weak, case-2 when link-1 is weak, and case-3 when neither link-1 nor link-2 is weak. First we consider case-3. We also  consider the general case when $\mcal{U}^{\stonen}(\alpha_{w}),\mcal{U}^{\sttwon}(\alpha_{w}),\mcal{U}^{\stthrn}(\alpha_{w})$ and $\mcal{U}^{\stnn}(\alpha_{w})$ are not empty domain-sets.  For  $z\in\{1...W-1\}$, it is easy to re-write $\tau_{1}(\alpha_{z},1,P^{\sttwon}_{1},P^{\stthrn}_{1})$ and $\tau_{2}(\alpha_{z},1,P^{\stonen}_{2},P^{\stthrn}_{2})$ from (\ref{eqn:S_fun}) in terms of $\tau_{1}(\alpha_{z},1,P^{\sttwon}_{1},1)$ and $\tau_{2}(\alpha_{z},1,P^{\stonen}_{2},0)$ as follows:
{\normalsize
\begin{eqnarray}\label{eqn:taut_easy}
	\begin{array}{lll}
		\tau_{1}(\alpha_{z},1,P^{\sttwon}_{1},P^{\stthrn}_{1})\hspace{-0.2cm}&=&\hspace{-0.2cm} \tau_{1}(\alpha_{z},1,P^{\sttwon}_{1},1)- P^{\stthrn}_{2}(\alpha_{z})\mcal{R}^{\stthrn}_{1}(\alpha_{z}),\\
					
		\tau_{2}(\alpha_{z},1,P^{\stonen}_{2},P^{\stthrn}_{2})\hspace{-0.2cm}&=&\hspace{-0.2cm} \tau_{2}(\alpha_{z},1,P^{\stonen}_{2},0)+ P^{\stthrn}_{2}(\alpha_{z})\mcal{R}^{\stthrn}_{2}(\alpha_{z}).
	\end{array}
\end{eqnarray}
}
Now after equating the inflow rate to that of outflow, i.e. $\tau_{1}(\alpha_{z},1,P^{\sttwon}_{1},P^{\stthrn}_{1})=\tau_{2}(\alpha_{z},1,P^{\stonen}_{2},P^{\stthrn}_{2})$, and solving for $P^{\stthrn}_{1}(\alpha_{z})$ or $P^{\stthrn}_{2}(\alpha_{z})$, we get:
\begin{eqnarray}\label{eqn:P_C_both}
	\begin{array}{lll}
	P^{\stthrn}_{2}(\alpha_{z})= \dfrac{\tau_{1}(\alpha_{z},1,P^{\sttwon}_{1},1)-\tau_{2}(\alpha_{z},1,P^{\stonen}_{2},0)}{\mcal{R}^{\stthrn}_{1}(\alpha_{z})+\mcal{R}^{\stthrn}_{2}(\alpha_{z})},\quad
	P^{\stthrn}_{1}(\alpha_{z})= \dfrac{\tau_{2}(\alpha_{z},1,P^{\stonen}_{2},1)-\tau_{1}(\alpha_{z},1,P^{\sttwon}_{1},0)}{\mcal{R}^{\stthrn}_{1}(\alpha_{z})+\mcal{R}^{\stthrn}_{2}(\alpha_{z})}.	
	\end{array}
\end{eqnarray}
After substituting (\ref{eqn:P_C_both}) in (\ref{eqn:taut_easy}), we get:
{\normalsize
	\begin{IEEEeqnarray}{rcl}\label{eqn:throu0}
		\begin{array}{lll}
		\hspace{-0.1cm}\label{eqn:sys_tau}\tau_{1}(\alpha_{z},1,P^{\sttwon}_{1},P^{\stthrn}_{1})&=&\tau_{2}(\alpha_{z},1,P^{\stonen}_{2},P^{\stthrn}_{2})= \alpha_{z}\tau_{1}(\alpha_{z},P^{\sttwon}_{1},1)+(1-\alpha_{z})\tau_{2}(\alpha_{z},P^{\stonen}_{2},0),\\
		\tau_{t}(\alpha_{z})&=& \alpha_{z}\tau_{1}(\alpha_{z},P^{\sttwon}_{1},1)+(1-\alpha_{z})\tau_{2}(\alpha_{z},P^{\stonen}_{2},0)+\tau_{3}(\alpha_{z},P^{\sttwon}_{1},P^{\stonen}_{2}).
		\end{array}
	\end{IEEEeqnarray}
}
After substituting the values of $\tau_{1}(\alpha_{z},P^{\sttwon}_{1},1),\,\tau_{2}(\alpha_{z},P^{\stonen}_{2},0)$ and $\tau_{3}(\alpha_{z},P^{\sttwon}_{1},P^{\stonen}_{2})$ from (\ref{eqn:S_fun}) in (\ref{eqn:throu0}) and some manipulations, we get the expression of link-rate $\tau_{i}$ for $i\in\{1,2,3\}$ as follows:
\begin{eqnarray}\label{eqn:buffered_direct_tau}
	\hspace{-.0cm}\label{eqn:buffered_tau_forward}\tau_{1}(\alpha_{z},1,P^{\sttwon}_{1},P^{\stthrn}_{1})\hspace{-.2cm}&=&\hspace{-.2cm}\alpha_{z}\mcal{R}^{\{1,\sttwon,\stthrn,\stnn\}}_{1}(\alpha_{z})+(1-\alpha_{z})\mcal{R}_{2}^{2}(\alpha_{z})-\alpha_{z}  \ovl{P}^{\sttwon}_{1} \mcal{R}^{\{\sttwon,\stnn\}}_{1}(\alpha_{z}) +(1-\alpha_{z})P^{\stonen}_{2} \mcal{R}^{\{\stonen,\stnn\}}_{2},\nonumber\\		
	\hspace{-.0cm}\label{eqn:buffered_tau_backward}\tau_{2}(\alpha_{z},1,P^{\stonen}_{2},P^{\stthrn}_{2})\hspace{-.2cm}&=&\hspace{-.2cm}\alpha_{z}\mcal{R}_{1}^{1}(\alpha_{z})+(1-\alpha_{z})\mcal{R}^{\{2,\stonen,\stthrn,\stnn\}}_{2}(\alpha_{z})+\alpha_{z}P^{\sttwon}_{1} \mcal{R}^{\{\sttwon,\stnn\}}_{1}(\alpha_{z})-(1-\alpha_{z})\ovl{P}^{\stonen}_{2} \mcal{R}^{\{\stonen,\stnn\}}_{2},\nonumber\\
	\hspace{-.0cm}\label{eqn:direct_tau_forward}\tau_{3}(\alpha_{z},P^{\sttwon}_{1},P^{\stonen}_{2})\hspace{-.2cm}&=&\hspace{-.2cm}\mcal{R}^{\{3,\stonen\}}_{3}(\alpha_{z})+\alpha_{z}  \ovl{P}^{\sttwon}_{1} \mcal{R}^{\{\sttwon,\stnn\}}_{1}(\alpha_{z}) -(1-\alpha_{z})P^{\stonen}_{2} \mcal{R}^{\{\stonen,\stnn\}}_{2}(\alpha_{z}),\nonumber\\
	\hspace{-.0cm}\label{eqn:direct_tau_backward}\hspace{-.2cm}&=&\hspace{-.2cm}\mcal{R}^{\{3,\sttwon\}}_{3}(\alpha_{z})-\alpha_{z}P^{\sttwon}_{1} \mcal{R}^{\{\sttwon,\stnn\}}_{1}(\alpha_{z})+(1-\alpha_{z})\ovl{P}^{\stonen}_{2} \mcal{R}^{\{\stonen,\stnn\}}_{2}(\alpha_{z}).\nonumber
\end{eqnarray}
Substituting the expressions of link-rates $\tau_{i}$ given above in $\tau_{t}=\tau_{1}+\tau_{3}=\tau_{2}+\tau_{3}$, we get the expression of optimum system throughput as follows:
{\normalsize
\begin{eqnarray}\label{eqn:sys_rate}
	\hspace{-.5cm}\tau_{t}(\alpha_{z})&=&\alpha_{z}\mcal{R}^{\{1,\sttwon,\stnn,\stthrn\}}_{1}(\alpha_{z})+(1-\alpha_{z})\mcal{R}_{2}^{2}(\alpha_{z})+ \mcal{R}^{\{3,\stonen\}}_{3}(\alpha_{z}),\label{eqn:total_tau_forward}\nonumber\\
	\hspace{-.5cm} &=&\alpha_{z}\mcal{R}_{1}^{1}(\alpha_{z})+(1-\alpha_{z})\mcal{R}^{\{2,\stonen,\stnn,\stthrn\}}_{2}(\alpha_{z})+\mcal{R}^{\{3,\sttwon\}}_{3}(\alpha_{z})\label{eqn:total_tau_backward}.
\end{eqnarray}
}
Further, in order to prove the minimum constraint, we write first the expression of  $\tau_{t}(\alpha_{z+1})$, $\tau_{t}(\alpha_{z})$, and $\tau_{t}(\alpha_{z-1})$ from (\ref{eqn:sys_rate}) as follows:
\begin{eqnarray}\label{eqn:sys_rate1}
	\begin{array}{lll}
		\tau_{t}(\alpha_{z+1})&=&\alpha_{z+1}\mcal{R}_{1}^{1}(\alpha_{z+1})+(1-\alpha_{z+1})\mcal{R}^{\{2,\stonen,\stnn,\stthrn\}}_{2}(\alpha_{z+1})+\mcal{R}^{\{3,\sttwon\}}_{3}(\alpha_{z+1}),\\ 
		\hspace{-.0cm}\tau_{t}(\alpha_{z})&=&\alpha_{z}\mcal{R}^{\{1,\sttwon,\stnn,\stthrn\}}_{1}(\alpha_{z})+(1-\alpha_{z})\mcal{R}_{2}^{2}(\alpha_{z})+ \mcal{R}^{\{3,\stonen\}}_{3}(\alpha_{z}),\\ 
		\hspace{-.0cm}&=&\alpha_{z}\mcal{R}_{1}^{1}(\alpha_{z})+(1-\alpha_{z})\mcal{R}^{\{2,\stonen,\stnn,\stthrn\}}_{2}(\alpha_{z})+\mcal{R}^{\{3,\sttwon\}}_{3}(\alpha_{z}),\\
		\hspace{-.0cm}\tau_{t}(\alpha_{z-1})&=&\alpha_{z-1}\mcal{R}^{\{1,\sttwon,\stnn,\stthrn\}}_{1}(\alpha_{z-1})+(1-\alpha_{z-1})\mcal{R}_{2}^{2}(\alpha_{z-1})+ \mcal{R}^{\{3,\stonen\}}_{3}(\alpha_{z-1}).
	\end{array}
\end{eqnarray}
After substracting $\tau_{t}(\alpha_{z})$ from $\tau_{t}(\alpha_{z+1})$ and $\tau_{t}(\alpha_{z-1})$ using (\ref{eqn:sys_rate1}) and applying the rate continuity property from (\ref{eqn:linkrate_cont}), we get:
\begin{eqnarray}\label{eqn:linkrate_diff1}
	\begin{array}{lll}
		\tau_{t}(\alpha_{z+1})-\tau_{t}(\alpha_{z})&=&(\alpha_{z+1}-\alpha_{z})(\mcal{R}_{1}^{1}(\alpha_{z+1})-\mcal{R}^{2}_{2}(\alpha_{z})),\\ 
		\hspace{-.0cm}\tau_{t}(\alpha_{z-1})-\tau_{t}(\alpha_{z})&=&(\alpha_{z-1}-\alpha_{z})(\mcal{R}^{1}_{1}(\alpha_{z})-\mcal{R}_{2}^{2}(\alpha_{z-1})).
	\end{array}
\end{eqnarray}
Now using the buffer-stability properties given in (\ref{eqn:buf_conds}), we conclude that for $\alpha_{z}$ to be optimum for buffer-stability, we  require that $\mcal{R}_{1}^{1}(\alpha_{z+1})\geq\mcal{R}^{2}_{2}(\alpha_{z})$ and $\mcal{R}^{1}_{1}(\alpha_{z})\leq \mcal{R}_{2}^{2}(\alpha_{z-1})$, which concludes the proof for $z\in\{1,2,...,W-1\}$ since $\tau_{t}(\alpha_{z+1})\geq\tau_{t}(\alpha_{z})$ and $\tau_{t}(\alpha_{z-1})\geq\tau_{t}(\alpha_{z}).$
\par Now, it is evident that the expression (\ref{eqn:sys_rate}) is not valid for case-1 and 2 as the buffer cannot be balanced with $z\in\{1,2,...,W-1\}$.  However,  as described by (\ref{eqn:state0}) for case-1 and using similar arguments for case-2, the system throughput can be obtained by substituting respectively $z=0$ and $z=W$ in the first and second equation of (\ref{eqn:sys_rate}) as follows:
\begin{eqnarray}\label{eqn:sys_rate2}
	\hspace{-.5cm}\tau_{t}(\alpha_{0})&=&\alpha_{0}\mcal{R}^{\{1,\sttwon,\stnn,\stthrn\}}_{1}(\alpha_{0})+(1-\alpha_{0})\mcal{R}_{2}^{2}(\alpha_{0})+ \mcal{R}^{\{3,\stonen\}}_{3}(\alpha_{0})=\mcal{R}_{2}^{2}(\alpha_{0})+ \mcal{R}^{\{3,\stonen\}}_{3}(\alpha_{0}),\label{eqn:total_tau_forward2}\nonumber\\
	\hspace{-.5cm} \tau_{t}(\alpha_{W})&=&\alpha_{W}\mcal{R}_{1}^{1}(\alpha_{W})+(1-\alpha_{W})\mcal{R}^{\{2,\stonen,\stnn,\stthrn\}}_{2}(\alpha_{W})+\mcal{R}^{\{3,\sttwon\}}_{3}(\alpha_{W})=\mcal{R}_{1}^{1}(\alpha_{W})+\mcal{R}^{\{3,\sttwon\}}_{3}(\alpha_{W})\label{eqn:total_tau_backward2}.
\end{eqnarray}
In order to prove the minimum constraint, we substitute respectively $z=W-1$ and $z=1$ in the first and second equation of (\ref{eqn:sys_rate}) as follows:
\begin{eqnarray}\label{eqn:sys_rate3}
	\hspace{-.5cm}\tau_{t}(\alpha_{W-1})&=&\alpha_{W-1}\mcal{R}^{\{1,\sttwon,\stnn,\stthrn\}}_{1}(\alpha_{W-1})+(1-\alpha_{W-1})\mcal{R}_{2}^{2}(\alpha_{W-1})+ \mcal{R}^{\{3,\stonen\}}_{3}(\alpha_{W-1}),\label{eqn:total_tau_forward3}\nonumber\\
	\hspace{-.5cm} \tau_{t}(\alpha_{1})&=&\alpha_{1}\mcal{R}_{1}^{1}(\alpha_{1})+(1-\alpha_{1})\mcal{R}^{\{2,\stonen,\stnn,\stthrn\}}_{2}(\alpha_{1})+\mcal{R}^{\{3,\sttwon\}}_{3}(\alpha_{1})\label{eqn:total_tau_backward3}.
\end{eqnarray}
After substracting $\tau_{t}(\alpha_{0})$ from $\tau_{t}(\alpha_{1})$ and $\tau_{t}(\alpha_{W-1})$ from $\tau_{t}(\alpha_{W})$ using (\ref{eqn:sys_rate2}) and (\ref{eqn:sys_rate3}) and applying the rate continuity property from (\ref{eqn:linkrate_cont}), we get:
\begin{eqnarray*}\label{eqn:linkrate_diff2}
	\begin{array}{lll}
		\tau_{t}(\alpha_{1})-\tau_{t}(\alpha_{0})&=&\alpha_{1}(\mcal{R}_{1}^{1}(\alpha_{1})-\mcal{R}^{2}_{2}(\alpha_{0})),\\ 
		\hspace{-.0cm}\tau_{t}(\alpha_{W-1})-\tau_{t}(\alpha_{W})&=&(\alpha_{W-1}-\alpha_{W})(\mcal{R}^{1}_{1}(\alpha_{W})-\mcal{R}_{2}^{2}(\alpha_{W-1})).
	\end{array}
\end{eqnarray*}
It is evident using (\ref{eqn:linkrate_diff1}) and (\ref{eqn:linkrate_diff2}) that $\tau_{t}(\alpha_{0})$ and $\tau_{t}(\alpha_{W})$ are indeed minimum for case-1 and case-2 respectively.$\blacksquare$
\section*{Appendix B}\vspace{-0.15cm}
\section*{Proof of Lemma-5}\vspace{-0.15cm}
In this appendix, we derive the expression for CCDFs $F_{\gamma_{1},\gamma_{3}}^{c}(y_{1},y_{3})$ and $F_{\gamma_{2}}^{c}(y_{2})$. The expression for $f_{\gamma_{2}}(y_{2})$ can be found by differentiating $F_{\gamma_{2}}^{c}(y_{2})$. It is obvious from (\ref{eqn:InsSNR}) that $F_{\gamma_{2}}^{c}(y_{2})$ can be expressed as follows:
\begin{eqnarray*}
	\begin{array}{lll}
		F_{\gamma_{2}}^{c}(y_{2}) &=&  \Pr\left\{\min\left(\gamma_{max},\frac{\mcal{\gamma}_{p}}{|g_{2}|^2}\right)|h_{2}|^2\geq y_{2}\right\}.
	\end{array}
\end{eqnarray*}
In order to evaluate the $F_{\gamma_{2}}^{c}(y_{2})$, we use the CCDF of the inverse channel. Let $G_{i}$ be the inverse of transmit SNR, which is defined as $G_{i}=\min\left(\gamma_{max},\frac{\mcal{\gamma}_{p}}{|g_{i}|^2}\right)^{-1}=\max\left(\frac{1}{\gamma_{max}},\frac{|g_{i}|^2}{\mcal{\gamma}_{p}}\right)$ for $i\in\{1,2,3\}$. We first express $F_{\gamma_{2}}^{c}(y_{2})$ in terms of CCDF of $G_{2}$, i.e. $F_{G_{2}}^{c}(x)$, as follows:
{\normalsize
	\begin{eqnarray}\label{eqn:FG2}
		\begin{array}{lll}
			F_{\gamma_{2}}^{c}(y_{2}) &=&  \E_{G_{2}}\left[\Pr\{|h_{2}|^2\geq y_{2}G_2\}\right]=\E_{G_2}\left[\exp\left(-\frac{y_{2}G_2}{\Omega_{2}}\right) \right]=\displaystyle\int\limits_{0}^{\infty}\exp(-\frac{y_{2}x}{\Omega_{2}})f_{G_{2}}(x)dx,\\
			&=& 1-\frac{y_{2}}{\Omega_{2}}\displaystyle\int\limits_{0}^{\infty}\exp(-\frac{y_{2}x}{\Omega_{2}})F_{G_{2}}^{c}(x)dx,
		\end{array}
	\end{eqnarray}
}
where the last line is obtained after performing integration by parts. Now we evaluate $F_{G_{i}}^{c}(x)$ as follows:
{\normalsize
	\begin{eqnarray*}
		\begin{array}{lll}
			F_{G_{i}}^{c}(g)&=&\Pr\{G_i \geq g\}=\Pr\left\{\max\left(\frac{\mathcal{\gamma}_{p}}{\gamma_{max}},{|g_{i}|^2}\right)\geq g\mathcal{\gamma}_{p}\right\},\\
			&=&\Pr\left\{\frac{1}{\gamma_{max}}\geq g,|g_{i}|^2\leq\frac{\mathcal{\gamma}_{p}}{\gamma_{max}}\right\}+\Pr\left\{|g_{2}|^2\geq g\gamma_{p},|g_{i}|^2\geq\frac{\mathcal{\gamma}_{p}}{\gamma_{max}}\right\},\\
		\end{array}
	\end{eqnarray*}
}
where the last line is obtained after expanding the $\max$ argument. After expanding the second term into $\frac{1}{\gamma_{max}}\geq g$ and $\frac{1}{\gamma_{max}}\leq g$  and some simplification, we get:
{\normalsize
	\begin{eqnarray*}
		\begin{array}{lll}
			F_{G_{i}}^{c}(g)&=&\Pr\left\{\frac{1}{\gamma_{max}}\geq g\right\}+\Pr\left\{\frac{1}{\gamma_{max}}\leq g,|g_{i}|^2\geq g\gamma_{p}\right\},\\
						&=& u\left(\frac{1}{\gamma_{max}}-g\right)+u\left(g-\frac{1}{\gamma_{max}}\right)\exp\left(-\frac{g\mathcal{\gamma}_{p}}{\Omega_{ip}}\right).
		\end{array}
	\end{eqnarray*}
}
%
Substituting the value of $F_{G_{2}}^{c}(x)$ in (\ref{eqn:FG2}), we get:
{\normalsize
	\begin{eqnarray*}\label{eqn:Tnewvalue}
	\begin{array}{lll}
	F_{\gamma_{2}}^{c}(y_{2}) &=& 1-\frac{y_{2}}{\Omega_{2}}\int\limits_{0}^{\infty}\exp(-\frac{y_{2}x}{\Omega_{2}})F_{G_{2}}^{c}(x)dx,\\
	
	&=& 1-\frac{y_{2}}{\Omega_{2}}\int\limits_{0}^{{1}/{\gamma_{max}}}\exp(-\frac{y_{2}x}{\Omega_{2}})dx-\frac{y_{2}}{\Omega_{2}}\int\limits_{{1}/{\gamma_{max}}}^{\infty}\exp\left(-\left(\frac{y_{2}}{\Omega_{2}}+\frac{\gamma_{p}}{\Omega_{2p}}\right)x\right)dx,  \\
	
	
	&=& \exp\left(-\frac{y_{2}}{\lambda_{2}}\right)\left[1-p_{2}\frac{y_{2}}{y_{2}+\mu_{2}} \right]= \exp\left(-\frac{y_{2}}{\lambda_{2}}\right)\left[1-p_{2}+p_{2} \frac{1}{1+\frac{y_{2}}{\mu_{2}}} \right].
	\end{array}
	\end{eqnarray*}
}
In a similar way,  the expression for $F_{\gamma_{1},\gamma_{3}}^{c}(y_{1},y_{3})$ can also be derived.$\blacksquare$
\bibliographystyle{ieeetr}
\bibliography{Thesis_Biblography_0_26_08_2018,Thesis_Biblography_1_26_08_2018,Thesis_Biblography_2_26_08_2018}
\end{document}